\documentclass[11pt,a4paper]{article}
\pdfoutput = 1
\usepackage{amsmath}
\usepackage{amsfonts}
\usepackage{jheppub}
\usepackage{caption}

\title{On the Relevance of the Thermal Scalar}

\author[a]{Thomas G. Mertens,}
\author[a]{Henri Verschelde}
\author[b,c,d]{and Valentin I. Zakharov}
\affiliation[a]{Ghent University, Department of Physics and Astronomy\\
Krijgslaan, 281-S9, 9000 Gent, Belgium}
\affiliation[b]{ITEP, B. Cheremushkinskaya 25, Moscow, 117218 Russia,}
\affiliation[c]{Moscow Inst Phys \& Technol, Dolgoprudny, Moscow Region, 141700 Russia,}
\affiliation[d]{School of Biomedicine, Far Eastern Federal University, Sukhanova str 8, 
Vladivostok 690950 Russia}

\emailAdd{thomas.mertens@ugent.be}
\emailAdd{henri.verschelde@ugent.be}
\emailAdd{vzakharov@itep.ru}
\abstract{We discuss near-Hagedorn string thermodynamics in general spacetimes using the formalism of the thermal scalar. Building upon earlier work by Horowitz and Polchinski, we relate several properties of the thermal scalar field theory (i.e. the stress tensor and $U(1)$ charge) to properties of the highly excited or near-Hagedorn string gas. We apply the formulas on several examples. We find the pressureless near-Hagedorn string gas in flat space and a non-vanishing (angular) string charge in $AdS_3$. We also find the thermal stress tensor for the highly excited string gas in Rindler space.}

\begin{document}

\maketitle

\section{Introduction}
The relevance and interpretation of the thermal scalar for string thermodynamics has been a source of confusion in the past (see e.g. \cite{Barbon:2004dd} and references therein). It has been known for quite some time that the critical behavior of a Hagedorn system can be reproduced by a random walk of the thermal scalar. This picture was made explicit in curved spacetimes in \cite{Kruczenski:2005pj}\cite{theory}. These results show that thermodynamical quantities such as the free energy and the entropy can be rewritten in terms of the thermal scalar. In this note we investigate whether other properties of the near-Hagedorn string gas (the energy-momentum tensor and the string charge) can be written in terms of the thermal scalar. This has been largely done by Horowitz and Polchinski in \cite{Horowitz:1997jc} for the high energy behavior of the energy-momentum tensor. Here we extend their analysis.\\
On a larger scale, we are interested in finding a suitable description to handle the long highly excited strings that critical string thermodynamics predicts. The thermal scalar is intimately related to this phase of matter though it is, in general, unclear precisely how. In this note we report on some modest progress in this direction.\\ 
This paper is organized as follows. \\
In section \ref{HP} we rederive the final result of Horowitz and Polchinski in detail. We will provide details and clarifications as we go along. Section \ref{nonstat} contains the extension to spacetimes with non-zero $G_{\tau i}$ metric components, necessary for computing components of the stress tensor with mixed (time-space) components and for treating stationary spacetimes. In section \ref{thermaver} we will extend the result to the canonical ensemble. Then in section \ref{stringcharge} we will provide the analogous result for the string charge. Section \ref{correl} contains some results on a specific class of correlators. In section \ref{second} we provide an interesting alternative derivation of the same results using a second quantized (Green function) formalism which illustrates some issues from a different perspective. We demonstrate the formulas in section \ref{examples} on several examples: flat space, the $AdS_3$ WZW model and Rindler spacetime. A summary of the results is presented in section \ref{summary} and the appendix contains some supplementary material.

\section{Derivation of the Horowitz-Polchinski result}
\label{HP}
In \cite{Horowitz:1997jc}, the authors consider long strings with self-interactions. This study was further analyzed in \cite{Damour:1999aw}. As a byproduct in their calculations, they obtain an expression for the energy-averaged stress tensor in terms of the thermal scalar. It is this result that we will focus on in this note. We already emphasize that we will only be interested in the non-self-interacting part of the stress tensor: interactions with a background are taken into account but no self-interactions between the stringy fluctuations. \\

\noindent The spacetime energy-momentum tensor of a single string in a general background is given by\footnote{The subscript $l$ denotes the Lorentzian signature tensor, whereas a subscript $e$ denotes the Euclidean signature tensor. Also, to avoid any confusion, in this paper we will write $0$ for the Lorentzian time index and $\tau$ for the Euclidean time index.}
\begin{equation}
\label{firsteq}
T^{\mu\nu}_{l}(\mathbf{x},t) = - \frac{2}{\sqrt{-G}}\frac{\delta S_l}{\delta G_{\mu\nu}(\mathbf{x},t)} = \frac{1}{2\pi\alpha'}\int_{\Sigma} d^{2}\sigma \sqrt{h} h^{ab}\partial_aX^{\mu}\partial_bX^{\nu}\frac{\delta^{(D)}((\mathbf{x},t) - X^{\rho}(\sigma,\tau))}{\sqrt{-G(\mathbf{x},t)}}.
\end{equation}
Note that from a worldsheet point of view this definition is problematic, since varying the metric does not guarantee a conformal background. So although for classical string theory this definition is useful, as soon as one considers the quantum theory we encounter conceptual problems. This is why we shall continue fully from the spacetime (field theory) point of view to rederive the result of \cite{Horowitz:1997jc}. \\
The backgrounds we have in mind have no temporal dependence ($\partial_0 G^{\mu\nu}=0$). To proceed, let us consider the Euclidean energy-momentum tensor (sourcing the Euclidean Einstein equations):
\begin{equation}
\label{euclstress}
T^{\mu\nu}_{e}(\mathbf{x},\tau) = - \frac{2}{\sqrt{G}}\frac{\delta S_e}{\delta G_{\mu\nu}(\mathbf{x},\tau)}. 
\end{equation}
Instead of using the worldsheet non-linear sigma model (as in equation (\ref{firsteq})), we focus on the spacetime action of all string modes: the string field theory action. Then we restrict this action to the sum of the non-interacting parts of the different string states, let us call the resulting action $S_e$.

\subsection{The stress tensor as a derivative of the Hamiltonian}
This free action $S_e$\footnote{Or better, its Lorentzian signature counterpart.} can be used to write down a corresponding (Lorentzian) Hamiltonian $H_l$. 
First we wish to establish a relationship between the above Euclidean stress tensor and the (non-interacting) string field theory Hamiltonian as
\begin{equation}
\label{question}
T^{\mu\nu}_{e}(\mathbf{x}) \stackrel{?}{=} - \frac{2}{\sqrt{G}}\frac{\delta H_l}{\delta G_{\mu\nu}(\mathbf{x})}. 
\end{equation}
This statement does \emph{not} hold as an operator identity in the full non-interacting string field theory. Luckily, we only need it in suitable quantum expectation values. \\

\noindent We start by focusing on the thermal ensemble at temperature $\beta$. \\
The Euclidean stress tensor, averaged over $\beta$ in time equals
\begin{equation}
\label{betaav}
T^{\mu\nu}_{e,\beta} = \frac{1}{\beta} \int_{0}^{\beta} d\tau T^{\mu\nu}_{e}(\mathbf{x},\tau) = \frac{1}{\beta} \int_{0}^{\beta} d\tau \frac{-2}{\sqrt{G}}\frac{\delta S_{e}}{\delta G_{\mu\nu}(\mathbf{x},\tau)}.
\end{equation}
This averaging over $\beta$ looks a bit funny, but it will turn out to be quite helpful.\footnote{For black hole spacetimes on the other hand, the Euclidean manifold and the thermal manifold are naturally identified and hence this integral over $\beta$ is not that strange in that case.} Note though that the spacetimes that we consider are static (or stationary), implying a time-independent stress tensor expectation value. So the averaging procedure is trivial in the end. \\
We are interested in computing
\begin{equation}
\text{Tr}\left(T^{\mu\nu}_{e,\beta}e^{-\beta H_l}\right),
\end{equation}
where we sum over the Fock space of free-string states. The Lorentzian Hamiltonian $H_l$ is a sum over the different field Hamiltonians present in the string spectrum. Due to the integral over $\tau$, we immediately obtain
\begin{equation}
\label{previous}
\int_{0}^{\beta} d\tau T^{\mu\nu}_{e}(\mathbf{x},\tau) = \frac{-2}{\sqrt{G}}\frac{\delta S_{\text{thermal}}}{\delta G_{\mu\nu}(\mathbf{x})},
\end{equation}
with $S_{\text{thermal}} = \int_{0}^{\beta}d\tau L_e$, the string field theory action on the thermal manifold, restricted to the quadratic parts (non-interacting). This action differs by $S_e$ only in the range of the temporal coordinate. Important to note is that this action does not contain thermal winding modes, since we only Wick-rotated the Lorentzian fields. \\
\noindent Note that in equation (\ref{previous}), we only vary $S_{\text{thermal}}$ with respect to time-independent metric variations. This result is elementary, consider for instance a general functional in two space dimensions:
\begin{equation}
F = \int dx \int dy \rho(x,y) L(x,y).
\end{equation}
Varying only with respect to $y$-independent configurations $\rho(x)$, we obtain
\begin{equation}
\frac{\delta F}{\delta \rho(x)} = \int dy L(x,y).
\end{equation}

\noindent In general we can write for the (non-interacting) multi-string partition function:
\begin{equation}
\label{feyn}
Z_{\text{mult}} = \text{Tr}\left(e^{-\beta H_l}\right) = \int \left[\mathcal{D}\phi^a\right]e^{-S_{\text{thermal}}}.
\end{equation}
This relation is obtained by writing down this equality for each string field separately. In this formula $\phi^a$ labels all the different string fields with suitable (anti)periodic boundary conditions around the thermal circle.\footnote{Again we emphasize that no winding modes around the thermal circle are included at this stage: only the string fields obtained by a Wick rotation of the theory.} As is well known, this identity also holds for operator insertions. After some juggling with the above formulas, we immediately obtain
\begin{align}
\text{Tr}\left(T^{\mu\nu}_{e,\beta}e^{-\beta H_l}\right) &= \int \left[\mathcal{D}\phi^a\right] T^{\mu\nu}_{e,\beta}e^{-S_{\text{thermal}}} \\
\label{second2}
&= \frac{2}{\beta \sqrt{G}}\frac{\delta }{\delta G_{\mu\nu}} \int \left[\mathcal{D}\phi^a\right]  e^{-S_{\text{thermal}}} \\
&= \text{Tr}\left(\frac{-2}{\sqrt{G}}\frac{\delta H_l}{\delta G_{\mu\nu}}e^{-\beta H_l}\right).
\end{align}

\noindent As we noted before, the temporal average is actually irrelevant and we obtain
\begin{equation}
\label{box}
\boxed{
\text{Tr}\left( T^{\mu\nu}_{e} e^{-\beta H_l}\right) = \text{Tr}\left(\frac{-2}{\sqrt{G}}\frac{\delta H_l}{\delta G_{\mu\nu}}e^{-\beta H_l}\right).}
\end{equation}
This equality holds for any positive value of $\beta$ (larger than $\beta_H$). We can hence inverse Laplace transform both sides of (\ref{box}) and we obtain
\begin{equation}
\boxed{
\text{Tr}\left( T^{\mu\nu}_e \delta(H_l - E)\right) = \text{Tr}\left(\frac{-2}{\sqrt{G}}\frac{\delta H_l}{\delta G_{\mu\nu}}\delta(H_l - E)\right),}
\end{equation}
and this provides the microcanonical equivalent to (\ref{box}). It is in this sense that we understand the equality (\ref{question}) between stress tensor and Hamiltonian functional derivative to the metric. \\

\noindent To make the above more concrete, we pause here and follow the same logic for a massless complex scalar field. Suppose the Lorentzian spectrum of string fluctuations on some manifold contains a massless complex scalar field. First we Wick-rotate the theory to write down the Euclidean stress tensor that we wish to consider. Explicitly:
\begin{align}
S_e &\propto \int_{-\infty}^{+\infty} d\tau \int dV \sqrt{G} G^{\mu\nu} \partial_{\mu}\phi \partial_{\nu} \phi^*, \\
T^{\mu\nu}_e &\propto \left(\partial^{\mu}\phi \partial^{\nu}\phi^* + (\mu \leftrightarrow \nu)\right) - \frac{1}{2}G^{\mu\nu}\partial^{\rho}\phi \partial_{\rho}\phi^*.
\end{align}
This stress tensor, as an operator in the canonical formalism, is averaged over Euclidean time $\beta$ and then inserted in the thermal trace as above:
\begin{align}
\label{result}
\text{Tr}\left(T^{\mu\nu}_{e,\beta}e^{-\beta H_l}\right) = \int_{\phi(\mathbf{x},\tau) = \phi(\mathbf{x},\tau+\beta)} \left[\mathcal{D}\phi\right]\left[\mathcal{D}\phi^*\right]T^{\mu\nu}_{e,\beta}e^{-\int_{0}^{\beta} d\tau \int dV \sqrt{G} G^{\mu\nu} \partial_{\mu}\phi \partial_{\nu} \phi^*}.
\end{align}
The thermal action written down in the exponent is equal to $S_e$, up to the different temporal integration. Then one applies the manipulations as above to extract the stress tensor in terms of a metric derivative of the Hamiltonian and the result follows. \\

\noindent The extensions of this to other higher spin fields is straightforward and we hope the concrete treatment of the scalar field provided some insight in this procedure: one does not need the concrete form of the action nor the stress tensor: we only need the fact that the action and stress tensor both are quadratic in the fields at the non-interacting level (allowing a full decoupling of all fields) and a time-independent background. \\
In string theory we should finally sum this quantity over all the string fields present in the Lorentzian spectrum. 

\subsection{Microcanonical stress tensor}
\label{microc}
Let us now compute this energy-momentum tensor when \emph{averaging} over all (Lorentzian) string states with fixed energy $E$, for very large $E$.\footnote{The concrete criterion for `very large' will follow shortly.} 
\begin{equation}
\label{micro}
\left\langle T^{\mu\nu}_{e}(\mathbf{x})\right\rangle_E = \frac{\text{Tr}\left[T^{\mu\nu}_{e}(\mathbf{x})\delta(H_l-E)\right]}{\text{Tr}\delta(H_l-E)},
\end{equation}
where the $E$ outside the expectation value denotes the averaging over states with energy $E$.
We can rewrite this as
\begin{align}
\left\langle T^{\mu\nu}_{e}(\mathbf{x})\right\rangle_E &= \frac{2}{\sqrt{G(\mathbf{x})}\text{Tr}\delta(H_l-E)}\text{Tr}\frac{\delta}{\delta G_{\mu\nu}(\mathbf{x})}\theta (E - H_l) \\
 &= \frac{2}{\sqrt{G(\mathbf{x})}\text{Tr}\delta(H_l-E)}\frac{\delta}{\delta G_{\mu\nu}(\mathbf{x})}\text{Tr}\theta (E - H_l),
\end{align}
where in the second line we used the Hellmann-Feynman theorem to extract the functional derivative out of the expectation value. 
Now we evaluate the traces in the high energy regime. Let us take as the density of string states
\begin{equation}
\label{dos}
\rho (E) \propto \frac{e^{\beta_H E}}{E^{D/2+1}},
\end{equation}
with $D$ the number of non-compact dimensions. Such a density of states is quite general and encompasses a wide class of Hagedorn systems \cite{Horowitz:1997jc}. The integral can be done\footnote{Only the large $E$ part is taken into consideration. The lower boundary of the integral is left arbitrary.}
\begin{equation}
\int^E d\tilde{E}\frac{e^{\beta_H \tilde{E}}}{\tilde{E}^{D/2+1}} = -(-\beta_H)^{D/2} \Gamma(-D/2,-\beta_H E)
\end{equation}
in terms of the incomplete Gamma function and for large $E$ it behaves as
\begin{equation}
\Gamma(-D/2,-\beta_H E) \propto (-\beta_H E)^{-D/2-1}e^{\beta_H E}.
\end{equation}
This term carries the large $E$ behavior of the integral and we obtain:
\begin{equation}
\int^E d\tilde{E}\frac{e^{\beta_H \tilde{E}}}{\tilde{E}^{D/2+1}} \approx \frac{e^{\beta_H E}}{\beta_H E^{D/2+1}}.
\end{equation}
Note that one still has a choice here: one can choose either one string state (thus considering the single-string energy-momentum tensor as was done in \cite{Horowitz:1997jc}) or the entire string gas. The only difference is in what we use for the density of states at energy $E$. The dominant (at large $E$) part of both of these is actually the same (up to irrelevant prefactors or factors of $E$ in the denominator for $D=0$). The interested reader is advised to take a closer look at section 3 of \cite{Bowick:1989us}, where the link between the single-string density of states and the multi-string density of states is made explicit. Moreover the argument they presented in making the link between $\rho_{\text{single}}$ and $\rho_{\text{multi}}$ is independent of the background spacetime and applies equally well to a general curved (stationary) spacetime. Further discussions (in flat space) are given in \cite{Barbon:2004dd}\cite{Deo:1989bv}. From the above calculation we see that both yield the same result:\footnote{Note that the precise proportionality constant in (\ref{dos}) depends on the background fields. But taking the functional derivative with respect to the metric of this term is always subdominant in the large $E$ limit. Hence, in the dominant contribution, the proportionality constant cancels immediately between numerator and denominator of (\ref{micro}).} 
\begin{equation}
\label{cancell}
\left\langle T^{\mu\nu}_{e}\right\rangle_E \approx \frac{2}{\sqrt{G}}e^{-\beta_H E}\frac{\delta}{\delta G_{\mu\nu}}\left(\frac{e^{\beta_H E}}{\beta_H}\right).
\end{equation}
Note that this is irrespective of the value of $D$, the number of non-compact dimensions. 
We rewrite this as
\begin{equation}
\label{equ1}
\left\langle T^{\mu\nu}_{e}\right\rangle_E \approx \frac{1}{\beta_H^{2}\sqrt{G}}\frac{\delta \beta_H^2}{\delta G_{\mu\nu}}\left( E - \frac{1}{\beta_H}\right) \approx \frac{E}{\beta_H^{2}\sqrt{G}}\frac{\delta \beta_H^2}{\delta G_{\mu\nu}}
\end{equation}
where we consider energies $E \gg \frac{1}{\beta_H}$, which is the large $E$ criterion alluded to at the start of this subsection. \\

\begin{figure}[h]
\centering
\includegraphics[width=0.5\textwidth]{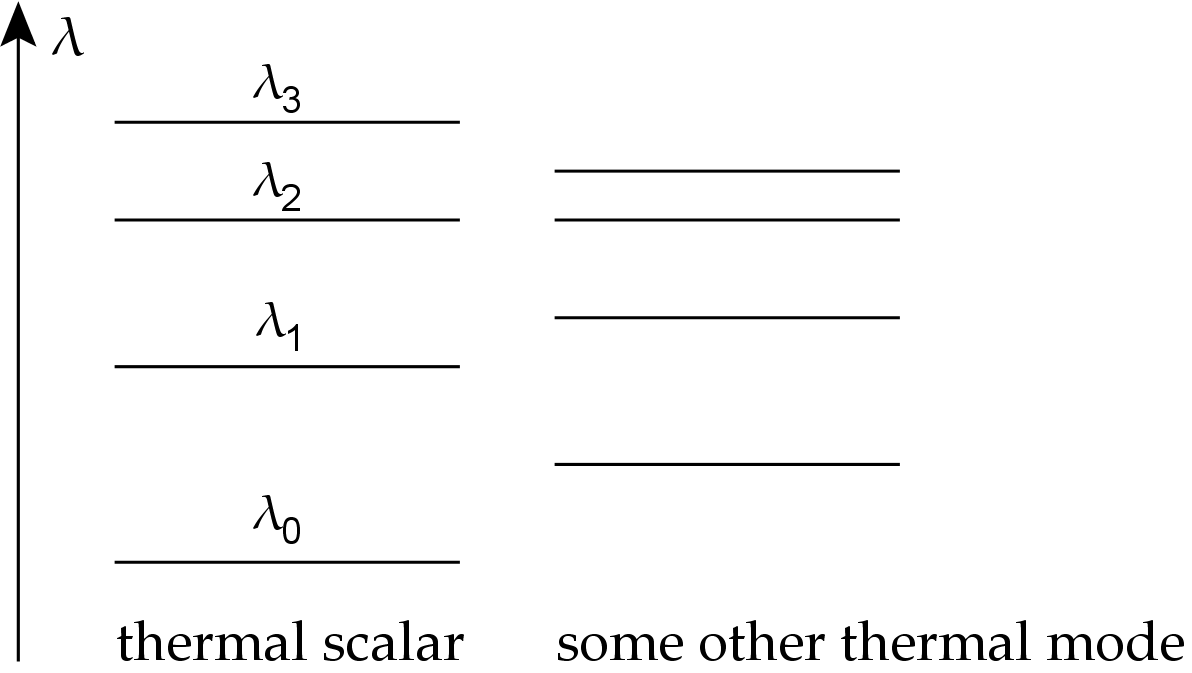}
\caption{Eigenvalue spectrum of the thermal scalar whose lowest eigenvalue $\lambda_0$ gives the dominant Hagedorn behavior of the thermal gas of strings. Higher eigenvalues are subdominant, even to some lower eigenvalues of other thermal modes (such as for instance the twice wound string state).}
\label{spectrum}
\end{figure}

\noindent At this point, it is instructive to recall some of the salient properties of the thermal scalar. This state is a singly wound state (around the thermal direction) on the thermal manifold that becomes massless precisely at $\beta = \beta_H$. At higher temperatures it becomes tachyonic. Just as above, we only consider the non-interacting gas so we only need the propagator part of the thermal scalar and no interactions with either itself or with other fluctuations. It is this propagator part that fully determines the Hagedorn temperature, by considering the eigenvalue problem associated to it. More in detail, suppose we are considering the thermal scalar partition function:
\begin{equation}
Z_{\text{th.sc.}} = \int \left[\mathcal{D}\phi\right]e^{-S_{\text{th.sc.}}},
\end{equation}
where $S_{\text{th.sc.}}$ contains only the quadratic parts of the field theory action of the thermal scalar. This action can be rewritten (after integration by parts) in the form:
\begin{equation}
S_{\text{th.sc.}} \sim \int dV e^{-2\Phi}\sqrt{G}T^* \hat{\mathcal{O}}T
\end{equation}
for some quadratic differential operator $\hat{\mathcal{O}}$. Then construct a complete set of eigenfunctions of $\hat{\mathcal{O}}$ as $\hat{\mathcal{O}}T_n = \lambda_n T_n$ (normalized in the canonical way). Then it is immediate that 
\begin{equation}
Z_{\text{th.sc.}} = \text{det}^{-1}\hat{\mathcal{O}},
\end{equation}
or in terms of the free energy $\beta F \approx - \text{ln} Z_{\text{th.sc.}}$:
\begin{equation}
\beta F \approx \text{Tr}\text{ln}\hat{\mathcal{O}}.
\end{equation}
The approximation arises in the above formulas because we are near the Hagedorn temperature where the full thermal ensemble can be approximated by only the thermal scalar. \\
For instance for a discrete spectrum of $\hat{\mathcal{O}}$, we obtain
\begin{equation}
\beta F \approx \sum_n\text{ln}\lambda_n
\end{equation}
and it is clear that if an eigenvalue approaches zero from above, it will dominate the free energy. This is the mode of the thermal scalar that is the important one for the dominant near-critical thermodynamics. This point is worth emphasizing in a slightly different way: string theory reduces near the Hagedorn temperature to the most dominant mode of the thermal scalar. Higher modes of the thermal scalar field theory are subdominant and can be just as subdominant as modes from thermal fields other than the thermal scalar. A cartoon of this (for a discrete spectrum) is shown in figure \ref{spectrum}.
For the remainder of this paper, our focus will be only on this lowest eigenmode and the dominant thermodynamic behavior it entails. \\

\noindent Let us remark that this formula can be rewritten using the Schwinger proper time trick as
\begin{equation}
\label{heatk}
\beta F = -\int_0^{+\infty}\frac{ds}{s}\text{Tr}e^{-s\hat{\mathcal{O}}}
\end{equation}
and it is this formula that is directly related to the worldsheet evaluation of the free energy by the identification $s=\tau_2$, the imaginary part of the torus modulus \cite{theory}. \\

\noindent The thermal scalar action determines the critical temperature $\beta_H$ by the condition that for this value of the temperature, the lowest eigenvalue $\lambda_0$ of the associated operator $\hat{\mathcal{O}}$ is precisely zero. The eigenmodes and -values of this operator are denoted as $T_n$ and $\lambda_n$.\footnote{At this point, we include the additional assumption that the spectrum of the thermal scalar wave equation is discrete. If this is not the case, one should integrate out the continuous quantum numbers and consider the resulting operator. The meaning of this statement can be given in the heat kernel language of equation (\ref{heatk}). Suppose we are interested in some operator $\hat{\mathcal{O}}$ that includes a continuous quantum number $k$. Then we can write schematically
\begin{equation}
\text{Tr}e^{-s \hat{\mathcal{O}}} = \sum_n \int dk \rho_n(k) e^{-s\lambda_n(k)} = \sum_n f(n,s) e^{-s \tilde{\lambda_n}}
\end{equation}
where in the final line we integrated over $k$, the continuous quantum number and we extracted the dominant large $s$ exponential factor. The reasoning then applies if the lowest eigenvalue is left unchanged: $\lambda_0 = \tilde{\lambda_0}$, and this for all metrics infinitesimally displaced from the metric of interest.}
A concrete form of $\hat{\mathcal{O}}$ will be given further on for the type II superstring, but for now we remain more general. \\
So we have $\lambda_0(G_{\mu\nu}, \Phi, \beta_H) = 0$, where this equality holds for all backgrounds by definition. $\beta_H$ is on its own still a function of the background fields. Taking the total functional derivative of 
\begin{equation}
\lambda_0(G_{\mu\nu}, \Phi, \beta_H(G_{\mu\nu}, \Phi)) = 0
\end{equation}
with respect to the metric, we obtain
\begin{equation}
\left.\frac{\delta \lambda_0}{\delta G_{\mu\nu}(\mathbf{x})}\right|_{\text{Total}} = 0 = \frac{\delta \lambda_0}{\delta G_{\mu\nu}(\mathbf{x})} + \left.\frac{\partial \lambda_0}{\partial \beta^2}\right|_{\beta = \beta_H}\frac{\delta \beta_H^2}{\delta G_{\mu\nu}(\mathbf{x})},
\end{equation}
which can be rewritten as
\begin{equation}
\label{chain}
\frac{\delta \beta_H^2}{\delta G_{\mu\nu}(\mathbf{x})} = - \frac{\frac{\delta \lambda_0}{\delta G_{\mu\nu}(\mathbf{x})}}{\left.\frac{\partial \lambda_0}{\partial \beta^2}\right|_{\beta = \beta_H}}.
\end{equation}
The variations of the eigenvalues are given by definition as:
\begin{equation}
\frac{\delta \lambda_n}{\delta G_{\mu\nu}(\mathbf{y})} = \frac{\delta}{\delta G_{\mu\nu}(\mathbf{y})}\int d\mathbf{x}e^{-2\Phi}\sqrt{G}T_n^* \hat{\mathcal{O}} T_n = \frac{\delta}{\delta G_{\mu\nu}(\mathbf{y})} S_{\text{th.sc.}}\left[T_n\right]
\end{equation}
where the $n^{th}$ eigenmode $T_n$ (of the thermal scalar eigenvalue equation) is used.\footnote{Note that this functional derivative is only with respect to the metric, the parameter $\beta_H$ is left unchanged. In writing these expressions, we have also assumed the eigenfunctions to be normalized as 
\begin{equation}
\int d\mathbf{x}e^{-2\Phi}\sqrt{G}T_n^* T_n =1,
\end{equation}
although this is not strictly required.} In the final line we used a partial integration to relate the operator $\hat{\mathcal{O}}$ (being the inverse propagator) to the action $S_{\text{th.sc.}}$ evaluated on the $n^{th}$ eigenfunction. \\ 
Since the denominator in (\ref{chain}) is not $\mathbf{x}$-dependent, after using these results in equation (\ref{equ1}) we arrive at the conclusion that
\begin{equation}
\left\langle T^{\mu\nu}_{e}\right\rangle_E \propto \left.T^{\mu\nu}_{\text{th.sc.}}\right|_{\text{on-shell}},
\end{equation}
where the meaning of the subscript \emph{on-shell} is that the stress tensor of the thermal scalar should be evaluated on the zero-mode wavefunction solution. This zero-mode wavefunction coincides with the classical field satisfying the Euler-Lagrange equations. Hence the \emph{classical} thermal scalar energy-momentum tensor determines the time-averaged high energy stress-momentum tensor of Lorentzian string states.

\subsection{Explicit form for type II superstrings}
One can be more explicit if we know the precise form of (the non-interacting part of) the thermal scalar action (including all its $\alpha'$-corrections). For the bosonic string, we discussed in several specific examples \cite{Mertens:2013zya}\cite{Mertens:2014nca} that this action gets corrections and we do not know for a general background what these look like. For type II superstrings however, the corrections appear not to be present. We present a heuristic argument in favor of this in appendix \ref{argu}. We hence focus on type II superstrings and, assuming there are no corrections for any background, we can compute the above energy-momentum tensor explicitly.\\
In particular the thermal scalar action for type II superstrings is given by
\begin{equation}
\label{thaction}
S_{\text{th.sc.}} = \int d^{d}x e^{-2\Phi}\sqrt{G}\left[G^{ij}\partial_i T \partial_j T^* + \frac{\beta^2G_{\tau\tau}-\beta_{H,\text{flat}}^2}{4\pi^2\alpha'^2}TT^*\right],
\end{equation}
where $G$ denotes the determinant of the metric (including $G_{\tau\tau}$) and $d$ the total number of spatial dimensions. Note that an overall factor of $\beta$ arising from the integration over $\tau$ has been dropped, and in fact we define the thermal scalar action with the above (canonical) normalization. This will prove to be the correct approach. Using the above action, one obtains
\begin{align}
\label{thenergy}
\frac{e^{2\Phi}}{\sqrt{G}}\frac{\delta \lambda_0}{\delta G_{\tau\tau}} &= \frac{\beta^2 TT^*}{4\pi^2\alpha'^2} + \frac{1}{2}G^{\tau\tau}\left[G^{ij}\partial_i T \partial_j T^* + \frac{\beta^2G_{\tau\tau}-\beta_{H,\text{flat}}^2}{4\pi^2\alpha'^2}TT^*\right], \\
\frac{e^{2\Phi}}{\sqrt{G}}\frac{\delta \lambda_0}{\delta G_{ij}} &= -\frac{\nabla^{i}T^* \nabla^{j} T + \nabla^{j}T^* \nabla^{i} T}{2} + \frac{1}{2}G^{ij}\left[G^{kl}\partial_k T \partial_l T^* + \frac{\beta^2G_{\tau\tau}-\beta_{H,\text{flat}}^2}{4\pi^2\alpha'^2}TT^*\right], \\
\frac{\partial \lambda_0}{\partial \beta^2} &= \int d^{d}x \sqrt{G} e^{-2\Phi} G_{\tau\tau}\frac{TT^*}{4\pi^2\alpha'^2}. 
\end{align}
The final line above is a fixed number, let us call it $N$. Integrating the time component of this energy-momentum tensor yields\footnote{
From here on, we focus on constant dilaton backgrounds, so $\Phi = \Phi_0$. In writing these expressions, we used the fact that the thermal scalar is a zero-mode: its on-shell action vanishes.} 
\begin{align}
-\int \left\langle T_{\tau,e}^{\tau}\right\rangle_E \sqrt{G} dV &= -\int G_{\tau\tau}\left\langle T^{\tau\tau}_{e}\right\rangle_E \sqrt{G} dV=  \frac{E}{\beta_H^{2}} \int \frac{\frac{G_{\tau\tau}}{\sqrt{G}}\frac{\delta \lambda_0}{\delta G_{\tau\tau}}}{\left.\frac{\partial \lambda_0}{\partial \beta^2}\right|_{\beta = \beta_H}} \sqrt{G} dV \\ 
&= \frac{E}{\beta_H^{2}} \frac{\int dV \sqrt{G} G_{\tau\tau}\frac{\beta_H^2 TT^*}{4\pi^2\alpha'^2}}{\left.\frac{\partial \lambda_0}{\partial \beta^2}\right|_{\beta = \beta_H}} = E, 
\end{align}
which means $-\left\langle T_{\tau,e}^{\tau}\right\rangle$ can be interpreted as the (Lorentzian signature) energy density. The precise equality between the quantum high-energy stress tensor and the classical thermal scalar stress tensor then becomes
\begin{equation}
\left\langle T^{\mu\nu}_{e}\right\rangle_E \approx \left.\frac{E}{2N \beta_H^2} T^{\mu\nu}_{\text{th.sc.}}\right|_{\text{on-shell}},
\end{equation}
where the classical on-shell thermal scalar energy-momentum tensor is computed using the thermal scalar wavefunction with an arbitrary normalization in principle,\footnote{The normalization cancels between the numerator $\left.T^{\mu\nu}_{\text{th.sc.}}\right|_{\text{on-shell}}$ and the denominator $N$.} although we shall assume it is normalized in the standard fashion.

\section{Extension to spacetimes with $G_{\tau i} \neq 0$}
\label{nonstat}
In the previous section we focused on $G_{\tau i} = 0$. There are two cases when one wants to extend this assumption. Firstly, if one wishes to know $T^{\tau i}$, we need to compute $\frac{\delta S_{\text{th.sc.}}}{\delta G_{\tau i}}$. Secondly, a stationary (non-static) spacetime (such as a string-corrected Kerr black hole) contains $G_{\tau i}$ metric components. Note that even the metric of static spacetimes may be rewritten in the reference frame of a moving observer which then possibly includes non-zero $G_{\tau i}$ components (such as flat space as seen by a rotating observer). \\
The entire derivation presented in the previous section does not explicitly use the fact that $G_{\tau i} = 0$, except in writing down the form of the thermal scalar action given in (\ref{thaction}). In a more general spacetime, the action to be used is:
\begin{equation}
S_{\text{th.sc.}} = \int d^{D}x \sqrt{G}e^{-2\Phi}\left(\bar{G}^{ij}\partial_i T \partial_j T^{*} + \frac{\beta^2 G_{\tau\tau} - \beta_{H,\text{flat}}^2}{4\pi^2\alpha'^2}TT^*\right),
\end{equation}
where $\bar{G}^{ij}$ is the matrix inverse of $G_{ij} - \frac{G_{\tau i}G_{\tau j}}{G_{\tau\tau}}$ as we derived in detail in \cite{theory}. 
In fact, a simple calculation shows that $\bar{G}^{ij} = G^{ij}$. It is interesting to note that the above effective spatial geometry (encoded in $\bar{G}^{ij}$) is precisely the spatial metric one would obtain when operationally defining distance in a stationary (non-static) gravitational field.\footnote{See for instance section \S 84 in the second volume of Landau and Lifshitz (Classical Field Theory).} The random walking particle (as described by the thermal scalar) is sensitive to this effective metric. It appears that T-duality in the (Euclidean) time direction encodes this information. \\
The only thing left to do is to compute $\frac{\delta \lambda_0}{\delta G_{\tau i}}$. For this, we need to determine $\frac{\delta \bar{G}^{ij}}{\delta G_{\tau k}}$. 
A simple calculation shows that\footnote{To compute this, one needs to be careful about varying with respect to a symmetric tensor: the factor of $2$ in the denominator is easily missed.}
\begin{equation}
\label{c1}
\frac{\delta \bar{G}^{ij}}{\delta G_{\tau k}} = - \frac{G^{ik} G^{\tau j} + (i \leftrightarrow j)}{2}.
\end{equation}
Using this result we obtain
\begin{align}
\frac{e^{2\Phi}}{\sqrt{G}}\frac{\delta \lambda_0}{\delta G_{\tau i}} &= -\left( G^{ki} G^{\tau l} + (k \leftrightarrow l)\right)\left(\frac{\partial_k T^* \partial_{l} T + \partial_{l}T^* \partial_{k} T}{4}\right) \nonumber \\
&+ \frac{1}{2}G^{\tau i}\left[G^{kl}\partial_k T \partial_l T^* + \frac{\beta^2G_{\tau\tau}-\beta_{H,\text{flat}}^2}{4\pi^2\alpha'^2}TT^*\right].
\end{align}
Importantly, if $G_{\tau i}$ vanishes for \emph{every} $i$, both terms are zero and $T^{\tau i}_{\text{th.sc.}} = 0$. This implies the highly excited string(s) does not carry any spatial momentum in a static spacetime, as we expect. \\

\noindent For completeness, we analogously obtain
\begin{equation}
\label{c2}
\frac{\delta \bar{G}^{ij}}{\delta G_{\tau \tau}} = - G^{i\tau} G^{j \tau},
\end{equation}
which should be used when computing $T^{\tau\tau}$ in a non-static, stationary spacetime (such as a Kerr black hole).\footnote{We remark that for Kerr black holes, one runs into trouble with the ergoregion where the asymptotically timelike Killing vector used to define thermodynamics becomes spacelike. We will not pursue a concrete example of stationary spacetimes in this paper.} \\

\noindent A consistency check can be performed by once more computing the total energy as a volume integral of a suitable stress tensor component. The computations are a bit more involved and we present the calculations in appendix \ref{stationary}. Suffice it to say that the results are consistent.

\section{Thermal average}
\label{thermaver}
As a slight modification, we study the thermal average of the time-averaged Euclidean energy-momentum tensor of a string gas:
\begin{equation}
\label{starting}
\left\langle T^{\mu\nu}_{e}\right\rangle_{\text{thermal}} = \frac{\text{Tr}\left(e^{-\beta H_l}T^{\mu\nu}_{e}\right)}{\text{Tr}\left(e^{-\beta H_l}\right)}
\end{equation}
in the near-Hagedorn regime $\beta \approx \beta_H$. We want to remark that it has been observed in the past that at high temperature, the canonical ensemble suffers from large fluctuations and one should resort to the microcanonical picture instead \cite{Mitchell:1987hr}\cite{Mitchell:1987th}\cite{Brandenberger:1988aj}\cite{Bowick:1989us}\cite{Deo:1989bv}\cite{Kutasov:2000jp}. Keeping in mind this issue, it is nevertheless instructive to proceed because we will see how the stress tensor of the thermal scalar contains information on the above expectation value. Moreover, at least naively, for theories with holographic duals one would expect the dual field theory to have a well-defined canonical ensemble, or alternatively one could be interested in more deeply analyzing how the discrepancy between both ensembles is realized both in the bulk and on the holographic boundary \cite{Berkooz:2000mz}. Hence we will proceed in the canonical ensemble.
Using (\ref{second2}), we can write
\begin{equation}
\label{starting2}
\frac{\text{Tr}\left(e^{-\beta H_l}T^{\mu\nu}_{e}\right)}{\text{Tr}\left(e^{-\beta H_l}\right)} = \frac{2}{\beta \sqrt{G} Z_{\text{mult}}}\frac{\delta }{\delta G_{\mu\nu}} Z_{\text{mult}},
\end{equation}
where $Z_{\text{mult}}$ is the multi-string partition function. Using $Z_{\text{mult}} = e^{-\beta F}$, we can also write
\begin{equation}
\label{expre}
\left\langle T^{\mu\nu}_{e}\right\rangle_{\text{thermal}} = - \frac{2}{\sqrt{G}}\frac{\delta F}{\delta G_{\mu\nu}}.
\end{equation}
In QFT, the expectation value of the energy-momentum tensor is badly divergent in curved spacetimes. The above formula relates this divergence to that of the free energy. But string theory does not contain the UV divergence of field theories. Therefore the stress tensor does not diverge in the UV.\\
In the near-Hagedorn limit, the thermal scalar provides the dominant contribution to the partition function in the sense that:
\begin{equation}
Z_{\text{mult}} \approx Z_{\text{th.sc.}} = \int \left[\mathcal{D}\phi\right]e^{-\beta S_{\text{th.sc.}}},
\end{equation}
where the thermal scalar field theory is used \cite{theory}. One then obtains 
\begin{align}
\label{ththerm}
\left\langle T^{\mu\nu}_{e}\right\rangle_{\text{thermal}} &\approx \frac{1}{\beta Z_{\text{th.sc.}}} \frac{2}{\sqrt{G}}\frac{\delta }{\delta G_{\mu\nu}}\int \left[\mathcal{D}\phi\right]e^{-\beta S_{\text{th.sc.}}} \\
&= \frac{1}{ Z_{\text{th.sc.}}}\int \left[\mathcal{D}\phi\right]\left(- \frac{2}{\sqrt{G}}\frac{\delta S_{\text{th.sc.}}}{\delta G_{\mu\nu}}\right)e^{-\beta S_{\text{th.sc.}}},
\end{align}
and indeed we see here that the normalization of the thermal scalar action we chose before (extracting a factor of $\beta$ from the action) is consistent. \\
Thus the energy-momentum tensor of the near-Hagedorn string gas (\ref{starting}) can be computed by looking at the one-loop expectation value of the thermal scalar energy-momentum tensor (\ref{ththerm}). 
From this procedure we see explicitly what the stress tensor of the thermal scalar encodes: it gives information on the time-averaged stress tensor of the near-Hagedorn string gas. 
\noindent We can proceed as follows. Let us start with (\ref{expre}). Assuming a density of high-energy single-string states as:
\begin{equation}
\rho_{\text{single}}(E) \propto \frac{e^{\beta_H E}}{E^{D/2+1}}
\end{equation}
we have
\begin{equation} 
\label{asyformF}
Z_1 = -\beta F = \text{Tr}_{\text{single}}\left(e^{-\beta H_l}\right) = \int^{+\infty}dE \rho(E)e^{-\beta E} \approx -C(\beta - \beta_H)^{D/2}\ln(\beta-\beta_H),
\end{equation}
where we trace over the single-string Hilbert space only. In writing this, we have used the Maxwell-Boltzmann approximation relating the single-string partition function $Z_1$ to the multi-string partition function $Z_{\text{mult}}$ through exponentiation. This equality is valid as soon as the singly wound string dominates on the thermal manifold, meaning for $\beta$ sufficiently close to $\beta_H$. The temperature-independent constant $C$ is present only for $D>0$ and contains the factors originating from integrating out continuous quantum numbers in $Z_1$. This expression holds for $D$ even. If $D$ is odd, the final logarithm should be deleted. The most singular term then yields\footnote{The constant $C$ does depend on the geometry, but the term $\frac{\delta C}{\delta G_{\mu\nu}} (\beta - \beta_H)^{D/2}\ln(\beta-\beta_H)$ is subdominant since $C$ does not depend on $\beta$.} 
\begin{equation}
\frac{\delta }{\delta G_{\mu\nu}} C(\beta - \beta_H)^{D/2}\ln(\beta-\beta_H) = -C \frac{D}{2}(\beta - \beta_H)^{D/2-1}\ln(\beta-\beta_H)\frac{\delta \beta_H}{\delta G_{\mu\nu}}.
\end{equation}
Collecting the results, we obtain
\begin{equation}
\left\langle T^{\mu\nu}_{e}\right\rangle_{\text{thermal}} \approx -\frac{C}{N}\frac{2}{\sqrt{G}}\left.\frac{\delta S_{\text{th.sc.}}}{\delta G_{\mu\nu}}\right|_{\text{on-shell}}\frac{(\beta - \beta_H)^{D/2-1}\ln(\beta-\beta_H)}{2\beta\beta_H}\frac{D}{2},
\end{equation}
which can be rewritten as
\begin{equation}
\label{mainresult}
\boxed{
\left\langle T^{\mu\nu}_{e}\right\rangle_{\text{thermal}} \approx \left. \frac{C}{N} T^{\mu\nu}_{\text{th.sc.}}\right|_{\text{on-shell}}\frac{(\beta - \beta_H)^{D/2-1}\ln(\beta-\beta_H)}{2\beta_H^2}\frac{D}{2}}
\end{equation}
and we see that we obtain the \emph{classical} thermal scalar energy-momentum tensor evaluated at $\beta = \beta_H$, multiplied by temperature-dependent factors that diverge (or are non-analytic) at $\beta = \beta_H$. The degree of divergence is related to that of the free energy by one $\beta$ derivative. Again the logarithm should be dropped when $D$ is odd.\footnote{For $D=0$ an exception occurs. Then the logarithm should be dropped as well and the $\frac{D}{2}$ factor should be deleted.}\\
As already noted above, this formula is derived in the canonical ensemble near the Hagedorn temperature. The large energy fluctuations of the canonical ensemble invalidate the equality between the microcanonical and canonical ensembles. This can be seen explicitly here by comparing the above result with formulas given in \cite{Brandenberger:2006xi}\cite{Brandenberger:2006vv} where a microcanonical approach is followed to determine the energy.\\
An interesting consistency check can be performed by integrating $G_{\mu \tau}\left\langle T^{\mu \tau}_{e}\right\rangle_{\text{thermal}}$ over the entire space. This yields\footnote{This check explicitly demonstrates that the result is consistent in the canonical ensemble, whereas a microcanonical approach would yield a different result. Note that
\begin{equation}
\int dV \sqrt{G} \left.{T_\tau^\tau}_{\text{th.sc.}}\right|_{\text{on-shell}} = -2 \beta_H^2 N.
\end{equation}
}
\begin{equation}
-\int \left\langle T^{\tau}_{\tau}\right\rangle_{\text{thermal}} \sqrt{G} dV = C\frac{D}{2}(\beta - \beta_H)^{D/2-1}\ln(\beta-\beta_H)
\end{equation}
which equals the internal energy $E$ of the thermodynamic system. This energy can also be determined as
\begin{equation}
E = \partial_{\beta}(\beta F) = \partial_{\beta} \left(C(\beta - \beta_H)^{D/2}\ln(\beta-\beta_H)\right) \approx C\frac{D}{2}(\beta - \beta_H)^{D/2-1}\ln(\beta-\beta_H)
\end{equation}
as it should be.\footnote{Note that in principle the above energy is formally negative below the Hagedorn temperature (except for $D=0$ \cite{Brandenberger:1988aj}), which is impossible. This is a typical feature of high-temperature string thermodynamics and points towards the fact that only including the most dominant contribution is not enough: one should include also subdominant contributions \cite{Deo:1988jj}\cite{Kutasov:2000jp}. Moreover, one should resort to the microcanonical ensemble instead.} 
We conclude that, much like the general expression for the free energy (\ref{asyformF}), the energy-momentum tensor can also be written down in a general background near its Hagedorn temperature (\ref{mainresult}). \\

\noindent The results of this section utilize the canonical ensemble, whereas the results in section \ref{HP} use the microcanonical ensemble. Qualitatively, both methods agree on the spatial distribution of the stress-energy.

\section{Extension to the String Charge}
\label{stringcharge}
We are interested in writing down an analogous formula for the string charge:
\begin{equation}
J^{\mu\nu}_{e} = - \frac{2}{\sqrt{G}}\frac{\delta S_e}{\delta B_{\mu\nu}}.
\end{equation}
Let us consider the high energy-averaged string charge and again focus on time-independent backgrounds: $\partial_\tau B_{\mu\nu} =0$, although we do allow temporal indices for $B$ here. \\
A closer look at the arguments presented in section \ref{HP} shows that the entire derivation can be copied to handle this case as well.
Hence upon replacing $G_{\mu\nu}$ with $B_{\mu\nu}$, we obtain an expression for the high-energy averaged string charge tensor as
\begin{equation}
\left\langle J^{\mu\nu}_{e}\right\rangle_E \approx  -\frac{E}{N\beta_H^2}\frac{1}{\sqrt{G}}\left.\frac{\delta S_{\text{th.sc.}}}{\delta B_{\mu\nu}}\right|_{\text{on-shell}}.
\end{equation} 
The expectation value of the string charge when averaged over high energy states is given by the classical charge tensor of the thermal scalar. \\
The thermal scalar itself, being a complex scalar field, has an additional $U(1)$ symmetry. It turns out that this $U(1)$ charge symmetry and the string charge of the thermal scalar are actually the same as we now show.\footnote{In fact this is readily shown as follows. Winding strings are charged under $B_{\tau i}$, hence it can be coupled to it through a conserved current. It is then no real surprise to find
\begin{equation}
\frac{\delta S_{\text{th.sc.}}}{\delta B_{\tau i}} \sim J^i
\end{equation}
with $J^i$ the conserved $U(1)$ current of the complex thermal scalar. It is interesting however to see it more explicitly.} The thermal scalar action is in general given by \cite{theory}:
\begin{align}
\label{wind}
&S_{\text{th.sc.}} \sim \int d^{D-1}x \sqrt{G}e^{-2\Phi} \nonumber \\
&\times\left(G'^{ij}\partial_{i}T\partial_{j}T^{*}+\frac{w^2\beta^2G'^{\tau\tau}}{4\pi^2\alpha'^2}TT^{*}+ G'^{\tau i}\frac{iw\beta}{2\pi\alpha'}\left(T\partial_{i}T^{*}- T^{*}\partial_{i}T\right)+ m^2TT^{*}\right),
\end{align}
where primes denote T-dual quantities. This action was written down in general, but we focus here on trivial dilatons: $\Phi = \Phi_0$. The $U(1)$ symmetry of this action leads to the following Noether current:
\begin{equation}
\label{noether}
J^{k} \propto G^{'ik}\left(T^{*}\partial_i T - T \partial_i T^{*}\right) + \frac{iw \beta}{\pi\alpha'}G^{'\tau k}TT^{*}.
\end{equation}
Explicitly, one can show that\footnote{These formulas hold also when $G_{\tau i} \neq 0$.}
\begin{align}
\frac{\partial G^{'\tau\tau}}{\partial B_{\tau k}} &= -2G^{'\tau k}, \\
\frac{\partial G^{'\tau i}}{\partial B_{\tau k}} &= -G^{'ik}, \\
\frac{\partial G^{'ij}}{\partial B_{\tau k}} &= 0.
\end{align}
Using these formulas to evaluate the expression 
\begin{equation}
\left.J^{\mu\nu}_{\text{th.sc.}}\right|_{\text{on-shell}} = -\frac{2}{\sqrt{G}}\left.\frac{\delta S_{\text{th.sc.}}}{\delta B_{\mu\nu}}\right|_{\text{on-shell}},
\end{equation}
we obtain that $\left.J^{ij}_{\text{th.sc.}}\right|_{\text{on-shell}} = 0$ and $\left.J^{\tau k}_{\text{th.sc.}}\right|_{\text{on-shell}} \propto \left.J^{k}\right|_{\text{on-shell}}$, which shows the equality between the $U(1)$ Noether current and the string charge of the thermal scalar. In terms of the highly excited string gas we are interested in, this leads to
\begin{align}
\left\langle J^{ij}_{e}\right\rangle_E &= 0 ,\\
\left\langle J^{\tau k}_{e}\right\rangle_E &\propto \left.J^{k}\right|_{\text{on-shell}}.
\end{align}
The Noether current of the thermal scalar determines the time-averaged, high-energy-averaged, string charge of a long string in a non-trivial background.\\

\noindent The surprising part of this analysis is that the high energy string average is sensitive to $B_{\mu\nu}$ even though each individual state (except for spatially wound states) is not (at the non-interacting level we consider here). \\

\noindent Finally we note that one can readily study this string charge also in the canonical ensemble (much like what was done in the previous section). One readily finds completely analogous expressions.

\section{Some Correlators}
\label{correl}
In this section we will be interested in studying a specific class of stress tensor correlators. The computations are more natural to perform in the canonical ensemble and we focus on this approach in this section. The canonical picture will allow us to make some specific clarifications on the factorization of the correlators as will be discussed further on.

\subsection{Correlators in the canonical ensemble}
\label{cancor}
We extend the analysis of section \ref{thermaver} to an energy-momentum correlator in the canonical ensemble:\footnote{Notice that a subscript $\beta$ has been written, denoting the averaging over Euclidean time as was considered earlier in equation (\ref{betaav}). Also, in principle this product of operators should be time-ordered, where the time-dependence of both of these operators is inside the temporal average. To avoid any more cluttering of the equations, we will not write this.}
\begin{align}
\left\langle T^{\mu\nu}_{e,\beta}(\mathbf{x}) T^{\rho\sigma}_{e,\beta}(\mathbf{y})\right\rangle_{\text{thermal}}.
\end{align}
Both energy-momentum tensors are individually time-averaged and evaluated at a spatial point $\mathbf{x}$ (or $\mathbf{y}$). Of course, such a correlator is not what we are interested in in the end, but it is at first sight the only type of correlator the methods described above can handle. This leads to the result that this correlator in the canonical ensemble is given by the same correlator of only the thermal scalar field theory ($Z_{\text{mult}} \approx Z_{\text{th.sc.}}$):
\begin{align}
\left\langle T^{\mu\nu}_{e,\beta}(\mathbf{x}) T^{\rho\sigma}_{e,\beta}(\mathbf{y})\right\rangle_{\text{thermal}} &\approx \frac{1}{\beta^2 Z_{\text{th.sc.}}} \frac{4}{\sqrt{G(\mathbf{x})}\sqrt{G(\mathbf{y})}}\frac{\delta }{\delta G_{\mu\nu}(\mathbf{x})}\frac{\delta }{\delta G_{\rho\sigma}(\mathbf{y})}\int \left[\mathcal{D}\phi\right]e^{-\beta S_{\text{th.sc.}}}. 
\end{align}
Throughout this section, we neglect contact terms. It seems hard to try to generalize this computation to more general types of correlators, since from the thermal scalar point of view, we are using the only way we can think of to compute its energy-momentum correlators. \\
To proceed, the same strategy as utilized before in section \ref{thermaver} can be followed. We write
\begin{equation}
\label{eq}
\frac{\frac{\delta^2 Z_{\text{th.sc.}}}{\delta G_{\mu\nu}\delta G_{\rho \sigma}}}{Z_{\text{th.sc.}}} = \frac{\delta^2 Z_1 }{\delta G_{\mu\nu} \delta G_{\rho\sigma}} + \frac{\delta Z_1}{\delta G_{\mu\nu}} \frac{\delta Z_1}{\delta G_{\rho\sigma}}.
\end{equation}
The second term represents the disconnected part of the correlator where the two stress tensors factorize. With the limiting behavior (\ref{asyformF}) in mind, we can compare the behaviors of both of these terms. We will focus here on fully compact spactimes (since these are the relevant ones for thermodynamics). For fully compact spaces, one has $Z_1 = - \ln(\beta-\beta_H)$ and one readily shows that both terms in (\ref{eq}) are equal. This then leads to 
\begin{equation}
\label{correla}
\left\langle T^{\mu\nu}_{e,\beta}(\mathbf{x}) T^{\rho\sigma}_{e,\beta}(\mathbf{y})\right\rangle \approx \frac{1}{2N^2\beta_H^{4}}\frac{1}{(\beta-\beta_H)^2}\left.T^{\mu\nu}_{\text{th.sc.}}(\mathbf{x})\right|_{\text{on-shell}}\left.T^{\rho\sigma}_{\text{th.sc.}}(\mathbf{y})\right|_{\text{on-shell}}.
\end{equation}
Such equalities can be generalized to more than two stress tensors and one finds in the near-Hagedorn limit that in fact all terms are proportional such that
\begin{equation}
\frac{\frac{\delta^n Z_{\text{th.sc.}}}{\delta G_{\mu\nu}\delta G_{\rho \sigma}\hdots}}{Z_{\text{th.sc.}}} = n! \frac{\delta Z_1}{\delta G_{\mu\nu}}\frac{\delta Z_1}{\delta G_{\rho\sigma}} \hdots.
\end{equation}
The prefactors of the analog of (\ref{eq}) in the expansion correspond precisely to the number of diagrams one can draw with a fixed number of disconnected components (each of which connecting a fixed number of points), as we will analyze more deeply in what follows. The main message here is that for a fully compact space, the correlator looks factorized because both connected and disconnected contributions become equal in the Hagedorn limit. \\
As a byproduct, the expectation value of $n$ stress tensors is given by:
\begin{equation}
\left\langle T^{\mu\nu}_{e,\beta} T^{\rho\sigma}_{e,\beta} \hdots \right\rangle \approx \frac{n!}{N^n}\frac{1}{2^n\beta_H^{2n}}\frac{1}{(\beta-\beta_H)^n}\left.T^{\mu\nu}_{\text{th.sc.}}\right|_{\text{on-shell}}\left.T^{\rho\sigma}_{\text{th.sc.}}\right|_{\text{on-shell}} \hdots 
\end{equation}
The analogous formulas for non-compact spacetimes will not be presented here.

\subsection{A puzzle on factorization}
The result that the stress tensor correlator (\ref{correla}) looks factorized may sound strange at first sight. After all, even in flat space one would expect (for the connected part) a behavior in terms of the distance between the two points. It is quite instructive to look into this matter in more detail. Actually, we do not need the added complication of the composite operators in the stress tensor to illustrate this, and so we concentrate on the free field propagator. We focus on flat space here and hope that the lesson we learn here will convince the reader that also in curved space, the near-Hagedorn limit taken above is consistent. \\
In flat Euclidean space, the Klein-Gordon propagator for a massive (complex) scalar field can be written down explicitly as
\begin{align}
\left\langle \phi(\mathbf{x}) \phi^*(\mathbf{0})\right\rangle = \frac{1}{(2\pi)^D}\int d^D \mathbf{q} \frac{e^{i \mathbf{q} \cdot \mathbf{x}}}{\mathbf{q}^2 + m^2} =  \left(\frac{1}{2\pi}\right)^{D/2}\left(\frac{m}{\left|\mathbf{x}\right|}\right)^{\frac{D-2}{2}}K_{\frac{D-2}{2}}(m\left|\mathbf{x}\right|).
\end{align}
In our case, we have $m \sim (\beta-\beta_H)^{1/2}$. We are interested in the massless limit and we only want the most dominant (non-analytic) contribution. Taylor expanding the modified Bessel function, one finds the series\footnote{Exactly the same caveats as before (below equation (\ref{asyformF})) apply here: the logarithm is only present for even $D$, except when $D=0$.}
\begin{align}
\label{seri}
&\left\langle \phi(\mathbf{x}) \phi^*(\mathbf{0})\right\rangle \approx \frac{C}{\left|\mathbf{x}\right|^{D-2}} + \sum_{i=0}^{+\infty} \frac{C_i m^{2i}}{\left|\mathbf{x}\right|^{D-2-2i}}+ C'm^{D-2}\ln m + \hdots
\end{align}
The first term is mass-independent (it becomes the UV cut-off in the coincident limit for a massless field), the second sum is regular as $m\to0$ and does not display non-analytic behavior. The dots denote terms of higher power in $m$ that result from multiplication of the third term by positive powers of $m$\footnote{And also a series in $m$ multiplied by $\text{ln}(\left|\mathbf{x}\right|)$.} and hence are less non-analytic than the third term. This term is hence the one we are interested in here and it does \emph{not} depend on the distance between the two points. It is straightforward to generalize this argument to the stress tensor correlator by including additional $\phi$ operator insertions and suitable derivatives. Also for the stress tensor itself, this propagator can be used (we will be more explicit in the next section). But one can already see that, upon setting $m \sim (\beta-\beta_H)^{1/2}$, the third term in this Taylor expansion has precisely the same temperature dependence as equation (\ref{mainresult}).\\

\noindent As as side remark, we note that in \cite{Horowitz:1997jc}, the authors write down the following correlator
\begin{equation}
\left\langle \phi\phi^*(\mathbf{x}) \phi\phi^*(\mathbf{0})\right\rangle \sim e^{-2m\left|\mathbf{x}\right|},
\end{equation}
which is also different from the above results: it is obtained by studying the asymptotic (large $\left|\mathbf{x}\right|$) behavior of the modified Bessel function (in the connected correlator). This equation is hence written down in the large $\left|\mathbf{x}\right|$ limit and only after that the $m\to0$ limit is taken. Our interest here lies in keeping $\left|\mathbf{x}\right|$ fixed while letting $m\to0$ (or $\beta \to \beta_H$): we focus on the regime $m\left|\mathbf{x}\right| \ll 1$. It is this change in limits that causes us to have a different final result. We further remark that this is the cause of the apparent violation of the cluster decomposition principle of the connected correlator:\footnote{We thank the anonymous referee for urging us to state this in more detail.} we are not taking the long distance limit, we are keeping the distance fixed and focus on the dominant (or non-analytic) parts of the correlator as the temperature reaches the Hagedorn temperature. To state this more explicitly, setting $m=0$ in (\ref{seri}), we obviously find
\begin{align}
\left\langle \phi(\mathbf{x}) \phi^*(\mathbf{0})\right\rangle \approx \frac{C}{\left|\mathbf{x}\right|^{D-2}},
\end{align}
where all other terms vanish. This is \emph{not} the procedure we are interested in here: we are discussing the process of letting $m$ go to zero and looking at the non-analytic behavior in $m$ during this procedure. All of this is related to the fact that, as was discussed in subsection \ref{microc}, we only focus on the lowest eigenmode of the thermal scalar and \emph{not} on the full scalar field theory to obtain the most dominant contribution to string thermodynamics in the Hagedorn regime. \\

\noindent A further way of appreciating these arguments, is to write the 2-point function (for a fully compact space) as 
\begin{equation}
\label{greenf}
\left\langle \phi(\mathbf{x}) \phi^*(\mathbf{\mathbf{y}})\right\rangle = \sum_n \frac{\psi_n(\mathbf{x})\psi_n^*(\mathbf{y})}{\lambda_n}
\end{equation}
and on taking the limit where $\lambda_0 \to 0$, this is approximated by
\begin{equation}
\left\langle \phi(\mathbf{x}) \phi^*(\mathbf{\mathbf{y}})\right\rangle \approx \frac{\psi_0(\mathbf{x})\psi_0^*(\mathbf{y})}{\lambda_0}.
\end{equation}
This makes it clear that a factorization is present. This type of argument (in terms of the Green function) will be made more explicit in the next section for the stress tensor and its correlators.
 
\section{Green function analysis}
\label{second}
In previous sections, we have found that the dominant contribution of the stress tensor in the canonical ensemble can be written in terms of the \emph{classical} stress tensor of the thermal scalar. It is instructive to study this from a Green function perspective as well. The stress tensor can be defined by a point-splitting procedure of the composite operator and related to a Green function. In this section we will analyze the stress tensor expectation value in the canonical ensemble again (using this point of view) and then continue our study of the correlators by elaborating on the comment made above around equation (\ref{greenf}). 

\subsection{Discrete spectrum}
For clarity, we focus first on purely compact spaces where the solution of $\hat{\mathcal{O}}\psi_n = \lambda_n \psi_n$ yields a discrete spectrum. Second quantization gives for the energy-momentum tensor of the thermal scalar (using point-splitting methods):
\begin{equation}
\label{operatorgf}
\left\langle \hat{T}^{ij}_{\text{th.sc.}}(\mathbf{x})\right\rangle = \lim_{\substack{\mathbf{x_1}\to \mathbf{x}\\ \mathbf{x_2}\to \mathbf{x}}} \left(\nabla_{1}^{i}\nabla_{2}^{j} + \nabla_{1}^{j}\nabla_{2}^{i} - G^{ij}\left[G^{kl}\nabla_{k}^{1}\nabla_{l}^{2} + \frac{\beta^2G_{\tau\tau}-\beta_{H,\text{flat}}^2}{4\pi^2\alpha'^2}\right]\right) G(\left.\mathbf{x_1}\right|\mathbf{x_2})
\end{equation}
where $G(\left.\mathbf{x_1}\right|\mathbf{x_2})$ is the scalar propagator of the thermal scalar given by
\begin{equation}
G(\left.\mathbf{x_1}\right|\mathbf{x_2}) = \sum_{n,m}\left\langle \mathbf{x_1}\right|\left.\psi_n\right\rangle\frac{\delta_{nm}}{\lambda_n}\left\langle \psi_{m}\right.\left|\mathbf{x_2}\right\rangle
\end{equation}
where the eigenfunctions and eigenvalues are found by solving:
\begin{equation}
\hat{\mathcal{O}}\psi_n(\mathbf{x}) = \left(-\nabla^2 - G^{ij}\frac{\partial_j G_{\tau\tau}}{G_{\tau\tau}}\partial_i + \frac{\beta^2G_{\tau\tau}-\beta_{H,\text{flat}}^2}{4\pi^2\alpha'^2} \right) \psi_n(\mathbf{x}) = \lambda_n \psi_n(\mathbf{x})
\end{equation}
and where $\nabla^2$ is the Laplacian on only the spatial submanifold. Likewise, the temporal component yields
\begin{equation}
\left\langle \hat{T}^{\tau\tau}_{\text{th.sc.}}(\mathbf{x})\right\rangle = \lim_{\substack{\mathbf{x_1}\to \mathbf{x}\\ \mathbf{x_2}\to \mathbf{x}}}\left(-\frac{\beta^2}{2\pi^2\alpha'^2} - G^{\tau\tau}\left[G^{kl}\nabla_{k}^{1}\nabla_{l}^{2} + \frac{\beta^2G_{\tau\tau}-\beta_{H,\text{flat}}^2}{4\pi^2\alpha'^2}\right]\right) G(\left.\mathbf{x_1}\right|\mathbf{x_2}).
\end{equation}
For a discrete spectrum and near the Hagedorn temperature, the propagator becomes:
\begin{equation}
G(\left.\mathbf{x_1}\right|\mathbf{x_2}) \approx \psi_0(\mathbf{x_1})\frac{1}{\lambda_0}\psi_{0}^*(\mathbf{x_2}).
\end{equation}
Since $\lambda_0 \sim \beta - \beta_H$, we obtain finally
\begin{align}
\left\langle \hat{T}^{ij}_{\text{th.sc.}}(\mathbf{x})\right\rangle &\approx \lim_{\substack{\mathbf{x_1}\to \mathbf{x}\\ \mathbf{x_2}\to \mathbf{x}}} \frac{\tilde{C}}{\beta-\beta_H}\left(\nabla_{1}^{i}\psi_0(\mathbf{x_1})\nabla_{2}^{j}\psi_{0}^*(\mathbf{x_2}) + \nabla_{1}^{j}\psi_0(\mathbf{x_1})\nabla_{2}^{i}\psi_{0}^*(\mathbf{x_2})\right. \\ \nonumber
&\left. \quad\quad\quad - G^{ij}\left[G^{kl}\nabla_{k}^{1}\psi_0(\mathbf{x_1})\nabla_{l}^{2}\psi_{0}^*(\mathbf{x_2}) + \frac{\beta^2G_{\tau\tau}-\beta_{H,\text{flat}}^2}{4\pi^2\alpha'^2}\psi_0(\mathbf{x_1})\psi_{0}^*(\mathbf{x_2})\right]\right) \\
&= \left.T^{ij}_{\text{th.sc.}}(\mathbf{x})\right|_{\text{on-shell}}\frac{\tilde{C}}{\beta-\beta_H}
\end{align}
where we obtain the \emph{classical} energy-momentum tensor of the thermal scalar. Analogously, we obtain
\begin{align}
\left\langle \hat{T}^{\tau\tau}_{\text{th.sc.}}(\mathbf{x})\right\rangle &\approx \lim_{\substack{\mathbf{x_1}\to \mathbf{x}\\ \mathbf{x_2}\to \mathbf{x}}} \frac{\tilde{C}}{\beta-\beta_H}\left( -\frac{\beta^2}{2\pi^2\alpha'^2}\psi_0(\mathbf{x_1})\psi_{0}^*(\mathbf{x_2})\right. \\ \nonumber 
&\left. \quad\quad\quad- G^{\tau\tau}\left[G^{kl}\nabla_{k}^{1}\psi_0(\mathbf{x_1})\nabla_{l}^{2}\psi_{0}^*(\mathbf{x_2}) + \frac{\beta^2G_{\tau\tau}-\beta_{H,\text{flat}}^2}{4\pi^2\alpha'^2}\psi_0(\mathbf{x_1})\psi_{0}^*(\mathbf{x_2})\right]\right) \\
&= \left.T^{\tau\tau}_{\text{th.sc.}}(\mathbf{x})\right|_{\text{on-shell}}\frac{\tilde{C}}{\beta-\beta_H}.
\end{align}
We have introduced here a constant $\tilde{C}$. These expression agree with those written down in equation (\ref{mainresult}).\footnote{Recall that for $D=0$, the logarithm and the $D/2$ factor should be dropped. The constant $\tilde{C}$ can then be fixed by comparing more closely.}\\

\begin{figure}[h]
\centering
\includegraphics[width=0.5\textwidth]{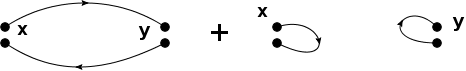}
\caption{Two possible Feynman diagrams for the (non-interacting) stress tensor correlator.}
\label{diagr}
\end{figure}

\noindent For correlators, we can schematically write\footnote{We have set $m_{local}^2 = \frac{\beta^2G_{\tau\tau}-\beta_{H,\text{flat}}^2}{4\pi^2\alpha'^2}$ representing the local mass term.} 
\begin{align}
\left\langle \hat{T}^{ij}_{\text{th.sc.}}(\mathbf{x}) \hat{T}^{kl}_{\text{th.sc.}}(\mathbf{y})\right\rangle &= \lim_{\substack{\mathbf{x_1}\to \mathbf{x}\\ \mathbf{x_2}\to \mathbf{x}}}\lim_{\substack{\mathbf{x_3}\to \mathbf{y}\\ \mathbf{x_4}\to \mathbf{y}}} \left\langle \left(\nabla^i \phi(\mathbf{x_1})\nabla^j \phi^*(\mathbf{x_2}) + \nabla^j \phi(\mathbf{x_1})\nabla^i \phi^*(\mathbf{x_2})\right.\right. \nonumber \\
&\quad\quad\quad\quad \left.- G^{ij}\left[G^{ab}\nabla_a \phi(\mathbf{x_1}) \nabla_b \phi^*(\mathbf{x_2}) + m_{local}^2 \phi(\mathbf{x_1})\phi^*(\mathbf{x_2})\right]\right) \nonumber \\
& \times\left(\nabla^k \phi(\mathbf{x_3})\nabla^l \phi^*(\mathbf{x_4}) + \nabla^l \phi(\mathbf{x_3})\nabla^k \phi^*(\mathbf{x_4}) \right. \nonumber \\ 
&\quad\quad\quad\quad \left.\left. - G^{kl}\left[G^{cd}\nabla_c \phi(\mathbf{x_3}) \nabla_d \phi^*(\mathbf{x_4}) + m_{local}^2 \phi(\mathbf{x_3})\phi^*(\mathbf{x_4})\right]\right)\right\rangle \\
&= \lim_{\substack{\mathbf{x_1}\to \mathbf{x}\\ \mathbf{x_2}\to \mathbf{x}}}\lim_{\substack{\mathbf{x_3}\to \mathbf{y}\\ \mathbf{x_4}\to \mathbf{y}}}\hat{\mathcal{D}}^{ijkl} \left\langle \phi(\mathbf{x_1})\phi^*(\mathbf{x_2})\phi(\mathbf{x_3})\phi^*(\mathbf{x_4})\right\rangle
\end{align}
with $\hat{\mathcal{D}}^{ijkl}$ a non-local differential operator. Next, we should contract this expectation value, which can only be done in two ways resulting in a connected diagram or a diagram consisting of two disconnected components since $\left\langle \phi \phi \right\rangle = \left\langle \phi^* \phi^* \right\rangle=0$ (figure \ref{diagr}).

\noindent Here again 
\begin{equation}
\left\langle \phi(\mathbf{x}) \phi^*(\mathbf{y}) \right\rangle \approx \frac{\psi_0 (\mathbf{x})\psi_0^*(\mathbf{y})}{\lambda_0}
\end{equation}
so we can just simply replace $\phi \to \psi_0$ in the correlator. The resulting expression is then twice\footnote{From the sum over all contractions of two $\phi$ and two $\phi^*$ vertices.} the product of the thermal scalar stress tensor:
\begin{equation}
\left\langle \hat{T}^{ij}_{\text{th.sc.}}(\mathbf{x}) \hat{T}^{kl}_{\text{th.sc.}}(\mathbf{y})\right\rangle \approx 2\left.T^{ij}_{\text{th.sc.}}(\mathbf{x})\right|_{\text{on-shell}} \left.T^{kl}_{\text{th.sc.}}(\mathbf{y})\right|_{\text{on-shell}} \frac{\tilde{C}^2}{(\beta-\beta_H)^2}.
\end{equation}
These results shed new light on the fact that we saw in subsection \ref{cancor} that for fully compact spaces the correlator is seen to factorize in the critical limit: as was briefly discussed there, all diagrams (both connected and disconnected) are equal in the Hagedorn limit. The combinatorics for multiple stress tensors yields a factor of $n!$ in agreement with our analysis in section \ref{cancor}. \\

\noindent It should not come as a big surprise to the reader that one can readily extend these formulas to the string charge and obtain very analogous results. 

\subsection{Continuous spectrum}
For a continuous spectrum, an analogous discussion can be made. Let us consider a continuous spectrum of states for the thermal scalar spectrum with the property that at $\beta = \beta_H$, the lowest eigenvalue of the associated eigenvalue problem vanishes.\footnote{As noted before, for a continuous spectrum this property is not trivial: it is possible that an integral over continuous quantum numbers shifts the critical eigenvalue to a non-zero value, as for instance happens in a linear dilaton background as is discussed elsewhere \cite{Mertens:2014cia}.} We write down
\begin{align}
\left\langle \hat{T}^{\mu\nu}_{\text{th.sc.}}\right\rangle = \lim_{\substack{\mathbf{x_1}\to \mathbf{x}\\ \mathbf{x_2}\to \mathbf{x}}} \hat{\mathcal{D}}^{\mu\nu}(\mathbf{x_1},\mathbf{x_2}) \int d k \rho(k)\frac{\psi_k(\mathbf{x_1})^{*}\psi_k(\mathbf{x_2})}{\lambda(k)}.
\end{align}
In this equation, $\hat{\mathcal{D}}^{\mu\nu}$ denotes the differential operator acting on the Green function to obtain the stress tensor (as written down in equation (\ref{operatorgf})) and $k$ labels the continuous quantum numbers with density of states $\rho(k)$. For a mixed continuous and discrete spectrum, one first focuses on the lowest discrete mode and then applies the method explained here. We are interested in the dominant limit as $\beta\to\beta_H$ or $\lambda(0) \to 0$. The dominant contribution to this limit can be found in the region of integration where $k \approx 0$. Hence we immediately extract the non-singular contributions:
\begin{align}
\left\langle \hat{T}^{\mu\nu}_{\text{th.sc.}}\right\rangle \approx \underbrace{\lim_{\substack{\mathbf{x_1}\to \mathbf{x}\\ \mathbf{x_2}\to \mathbf{x}}} \hat{\mathcal{D}}^{\mu\nu}(\mathbf{x_1},\mathbf{x_2}) \psi_0(\mathbf{x_1})^{*}\psi_0(\mathbf{x_2})}_{\left.T^{\mu\nu}_{\text{th.sc.}}\right|_{\text{on-shell}}} \int d k\frac{\rho(k)}{\lambda(k)}.
\end{align}
To derive the dominant contribution to this final integral in general, it is convenient to recall the form of the free energy in the critical regime in terms of the eigenvalues of the thermal scalar wave equation:
\begin{equation}
\label{todiff}
Z_1 = - \beta F \approx - \int d k \rho(k) \ln(\lambda(k)) \approx -C(\beta-\beta_H)^{D/2}\ln(\beta-\beta_H),
\end{equation}
where the same eigenvalues $\lambda(k)$ are used. This eigenvalue is generically such that $\lambda(0) \sim \beta-\beta_H$ and hence we can Taylor-expand:
\begin{equation}
\lambda(k) \approx B(\beta-\beta_H) + f(\beta,\beta_H) k + \mathcal{O}(k^2),
\end{equation}
for some proportionality constant $B$.
Differentiating (\ref{todiff}) with respect to $\beta$, we obtain\footnote{The contributions where $\rho(\mathbf{n})$ is differentiated is subdominant in the $\beta \approx \beta_H$ limit.}
\begin{equation}
- \int d k\frac{\rho(k)}{\lambda(k)}\left(B + \frac{\partial f}{\partial \beta} k + \mathcal{O}(k^2)\right) \approx
- \int d k\frac{B \rho(k)}{\lambda(k)} \approx -C\frac{D}{2}(\beta-\beta_H)^{D/2-1}\ln(\beta-\beta_H).
\end{equation}
Hence in the end we obtain
\begin{align}
\left\langle \hat{T}^{\mu\nu}_{\text{th.sc.}}\right\rangle \approx \left.T^{\mu\nu}_{\text{th.sc.}}\right|_{\text{on-shell}} \tilde{C}\frac{D}{2}(\beta-\beta_H)^{D/2-1}\ln(\beta-\beta_H),
\end{align}
for some constant $\tilde{C}$ and this is precisely the form we found in section \ref{thermaver}.\footnote{This perspective shows that, at least in the canonical ensemble, as soon as the continuous quantum numbers do not affect the critical temperature of the space we are interested in, the formulas discovered previously should be valid. In particular this implies that the rather awkward assumption made in section \ref{HP} about continuous spectra (integrating out continuous quantum numbers does not affect $\lambda_0$, also for variations on the background) seems in this case easier to handle.}

\section{Examples}
\label{examples}
The general strategy to proceed is then as follows. 
\begin{itemize}
\item[I.] Determine the thermal spectrum and identify the thermal scalar mode in the spectrum. This part was not discussed in this work and we will not focus on it here.
\item[II.] Compute the classical stress tensor of this mode. This then gives the dominant contribution to the stress tensor of the highly excited or near-Hagedorn string gas. Analogously for the string charge tensor.
\end{itemize}
Let us apply these results to some specific examples.
\subsection{Flat space}
Flat space in principle has a continuous spectrum of the thermal scalar eigenvalue problem. Instead of working with this continuum, we here take a more pragmatic road and simply introduce a finite-volume regularization (with periodic boundary conditions). This leads to the (normalized) thermal scalar wavefunction:
\begin{equation}
T_0 = \frac{1}{\sqrt{V}},
\end{equation}
which gives for the time-averaged, high-energy-averaged stress tensor:
\begin{align}
\left\langle T^{\tau\tau}_{e}(\mathbf{x})\right\rangle_E &\approx -\frac{E}{V}, \\
\left\langle T^{ij}_{e}(\mathbf{x})\right\rangle_E &\approx 0,
\end{align}
and indeed one expects the spacetime energy density to be $E/V$. The pressure vanishes in the high energy regime: thus high energy string(s) behave as a pressureless fluid (or as dust). Of course, what this means is that one should look at the next most dominant contribution. Nonetheless, the pressure is subleading with respect to other thermodynamic quantities at high energies.\\
Incorporating again the extra temperature factors to obtain the canonical average of the energy-momentum tensor, it is clear that the pressure still vanishes at $\beta \approx \beta_H$. It is interesting to learn that the Hagedorn string gas in flat space has vanishing pressure, very reminiscent of the final configuration of D-brane decay (see e.g. \cite{Sen:2004nf} and references therein). This fact was observed in the microcanonical approach of string gas cosmology as well \cite{Bassett:2003ck}\cite{Takamizu:2006sy}: the Hagedorn phase in flat space behaves as a pressureless fluid. The fact that our computation yields the same result provides confidence to our approach.\\

\noindent We also remark that choosing an \emph{arbitrary} compact spatial manifold (such as a compact Calabi-Yau) also leads to a pressureless state of matter. The reason is that such spatial manifolds have as their lowest thermal scalar eigenfunction the constant mode. A constant thermal scalar immediately leads to a vanishing $T^{ij}_{\text{th.sc.}}(\mathbf{x})$.

\subsection{$AdS_3$ space}
Let us look at the time-averaged, high energy string charge tensor in the WZW $AdS_3$ spacetime \cite{Gawedzki:1991yu}\cite{Maldacena:2000hw}\cite{Berkooz:2007fe}\cite{Lin:2007gi}. We will not give a self-contained treatment of this model here and the reader is refered elsewhere for details concerning the thermal spectrum.\\
A first subtlety here is again the fact that the thermal scalar is part of a band of continuous states. Just like in the flat space case, no corrections are generated by integrating the heat kernel over continuous quantum numbers \cite{Mertens:2014nca}. Assuming this property also holds for small variations in the $AdS_3$ metric and Kalb-Ramond background (or alternatively, working in the canonical ensemble where such a technicality does not present itself), we can proceed.\\
A second subtlety is that the density of high energy states includes a periodic part in $E$ \cite{Lin:2007gi}. A moment's thought shows that such periodic parts do not alter any of the conclusions.\footnote{More in detail, the density of single string states takes the schematic form
\begin{equation}
\rho_{\text{single}}(E) \sim E^p \frac{e^{\beta_H E}}{\left|\sin(a E)\right|}
\end{equation}
for some numbers $p$ and $a$ that we do not want to specify. In principle, an integral such as 
\begin{equation}
\int^{E}d\tilde{E} \tilde{E}^p \frac{e^{\beta_H \tilde{E}}}{\left|\sin(a \tilde{E})\right|}
\end{equation}
is infinite due to the extra poles caused by the sine factor. It is known that these are caused by the infinite $AdS_3$ volume felt by the so-called long strings. The difference with flat space is that this infinity does not cleanly factorize. Nonetheless, we are not interested in these divergences, but in the Hagedorn divergence. We hence obtain
\begin{equation}
\int^{E}d\tilde{E} \tilde{E}^p \frac{e^{\beta_H \tilde{E}}}{\left|\sin(a \tilde{E})\right|} \approx E^p \frac{e^{\beta_H E}}{\beta_H\left|\sin(a E)\right|}.
\end{equation}
The extra sine factor generated eventually cancels in the ratio of the thermal expectation values in (\ref{micro}), much like the $E^{D/2+1}$ factor did earlier on in equation (\ref{cancell}).} \\ 
A third subtlety is that the dominant thermal state is not just a single state, but is in fact all states labeled by an integer $q$ with $w=\pm1$ \cite{Mertens:2014nca}.\footnote{This is actually the reason for the appearance of the periodic factor in the high-energy density of states.} The $q$ quantum number corresponds to discrete momentum around the spatial cigar subspace. We choose $q=0$ as the thermal scalar state that has the information on the Hagedorn temperature.\footnote{We expect in general that the $q\neq0$ states will not contain the information on the Hagedorn temperature for infinitesimal metric variations. Note though, that we are only interested in how the critical temperature varies with the metric, for which the $q=0$ state is sufficient without caring about the $q\neq0$ states. We therefore pick $q=0$ and the situation is reduced to that discussed above. This issue is related to the fact that we only use $w=1$ and not both $w=\pm1$: one can choose between these for the critical temperature, and the answer is the same. In any case, the $AdS_3$ WZW model is quite pathological but we hope our treatment here will show how to handle other (more well-behaved) models.} \\

\noindent The background Kalb-Ramond field causes the string charge not to vanish. This can be seen from equation (\ref{noether}), where a nonvanishing $G'^{\tau k}$ is caused by turning on a Kalb-Ramond background. For $AdS_3$, we have $G'^{\tau\phi} = i$. Hence this yields a contribution to the string charge as
\begin{equation}
\left\langle J^{\tau \phi}_{e}(\mathbf{x})\right\rangle_E \propto \left.J^{\phi}(\mathbf{x})\right|_{\text{on-shell}} \propto \frac{\beta_H}{\alpha' \pi}T_0(\mathbf{x}) T^{*}_0(\mathbf{x}),
\end{equation}
where the wavefunction $T_0$ belongs to a continuous representation of the $SL(2,\mathbb{R})$ symmetry group and hence is not confined to the $AdS$ origin. Its precise form will not be written down here. What we emphasize in this context is that the charge does not vanish. The charge is directed alongside the angular cigar direction.

\subsection{Rindler space}
Let us finally apply these methods to Rindler space. We consider highly excited strings according to the Rindler observer. \\
The thermal metric is 
\begin{equation}
ds^2 = \frac{\rho^2}{\alpha'}d\tau^2 + d\rho^2 + \hdots
\end{equation}
where $\tau \sim \tau + 2 \pi \sqrt{\alpha'}$, the Rindler temperature. As it stands, we have used string-normalization for the metric. \\
We have shown in \cite{Mertens:2013zya} (see also \cite{Giveon:2012kp}\cite{Sugawara:2012ag}\cite{Giveon:2013ica}\cite{Giveon:2014hfa}) that the Hagedorn temperature in this space equals the Rindler temperature: $\beta_H = 2 \pi \sqrt{\alpha'}$ for type II superstrings. The thermal scalar was found to have a wavefunction of the form:
\begin{equation}
T_0(\rho) \sim e^{-\frac{\rho^2}{2\alpha'}}
\end{equation}
localized to the Rindler origin, or in terms of the black hole of which this is the near-horizon limit, localized to the black hole horizon.\\
In field theory, it is known that the Rindler observer (accelerated observer or fiducial observer) constructs his Fock space using his definition of positive-frequency modes. Moreover the vacuum constructed by an inertial observer (the \emph{Minkowski} vacuum) is seen by the Rindler observer as a thermal ensemble in terms of his coordinates and Fock space. The \emph{Rindler} vacuum is the vacuum the Rindler observer himself defines. Here we first focus on a high energy string(s) constructed on top of the Rindler vacuum.\\
In the end we will make a few remarks concerning the thermal ensemble itself (which upon including the Casimir contribution should coincide with the Minkowski stress tensor):
\begin{equation}
\left\langle T^{\mu\nu}\right\rangle_{M} = \text{Tr}_{R}\left(T^{\mu\nu}e^{-\beta H}\right).
\end{equation}
We are interested in the temperature-dependent part and hence the Casimir part does not interest us that much. In fact, for observations made by accelerated observers, the Casimir contribution is not detectable: it is only relevant when looking at the backreaction of the string(s) in the semi-classical Einstein equations. \\

\begin{figure}[h]
\centering
\begin{minipage}{.3\textwidth}
  \centering
  \includegraphics[width=\linewidth]{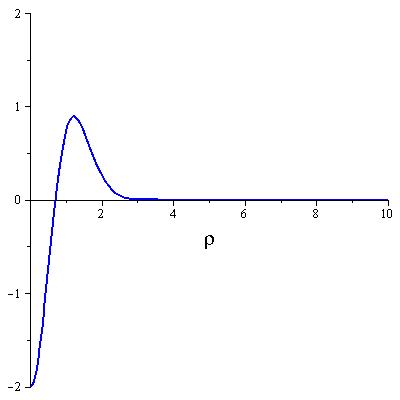}
  \caption*{(a)}
\end{minipage}
\begin{minipage}{.3\textwidth}
  \centering
  \includegraphics[width=\linewidth]{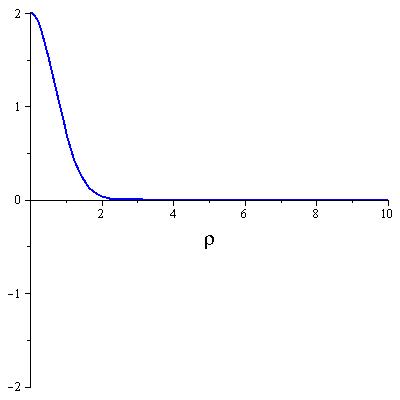}
  \caption*{(b)}
\end{minipage}
\begin{minipage}{.3\textwidth}
  \centering
  \includegraphics[width=\linewidth]{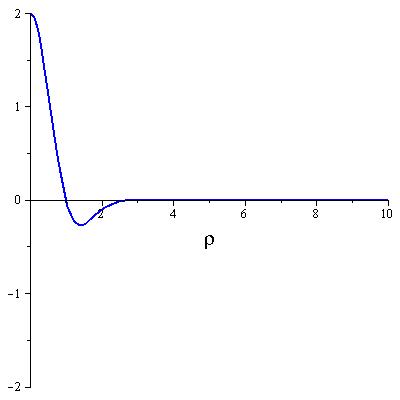}
  \caption*{(c)}
\end{minipage}
\caption{(a) Energy density $-\left\langle T^{\tau}_{\tau}(\mathbf{x})\right\rangle$ as a function of radial distance $\rho$ in units where $\alpha'=1$. (b) Radial pressure $\left\langle T^{\rho}_{\rho}(\mathbf{x})\right\rangle$. (c) Transverse pressure $\left\langle T^{i}_{i}(\mathbf{x})\right\rangle$.}
\label{form}
\end{figure}

\noindent These highly excited long strongs surrounding the event horizon are candidates for a microscopic description of the black hole membrane \cite{Susskind:1993ws}\cite{Kutasov:2005rr} thus it seems worthwhile to look into only this piece. We study the type II superstring and find the following quantities:
\begin{align}
\frac{e^{2\Phi}}{\sqrt{G}}\frac{\delta \lambda_0}{\delta G_{\tau\tau}} &= \frac{\beta_H^2}{4\pi^2\alpha'^2}e^{-\rho^2/\alpha'} + \frac{\alpha'}{2\rho^2}\left[\frac{\rho^2}{\alpha'^2} + \frac{\beta_H^2 \frac{\rho^2}{\alpha'}-\beta_{H,\text{flat}}^2}{4\pi^2\alpha'^2}\right]e^{-\rho^2/\alpha'}, \\
\frac{\partial \lambda_0}{\partial \beta^2} &= \frac{1}{4\pi^2\alpha'^2}\int_{0}^{+\infty}d\rho \left(\frac{\rho}{\sqrt{\alpha'}}\right)^3e^{-\rho^2/\alpha'} = \frac{1}{4\pi^2\alpha'^2}\frac{\sqrt{\alpha'}}{2}.
\end{align}
From this, one finds that
\begin{align}
\left\langle T^{\tau\tau}_{e}(\mathbf{x})\right\rangle_E &= -\frac{2E}{\sqrt{\alpha'}}\left(2-\frac{\alpha'}{\rho^2}\right)e^{-\rho^2/\alpha'},
\end{align}
and we readily check explicitly that indeed
\begin{equation}
-\int \left\langle T^{\tau}_{\tau,e}\right\rangle_E \sqrt{G} d^{D-1}x = \frac{2E}{\sqrt{\alpha'}}\int_{0}^{+\infty}d\rho \left(\frac{\rho}{\sqrt{\alpha'}}\right)^3\left(2-\frac{\alpha'}{\rho^2}\right)e^{-\rho^2/\alpha'} = E.
\end{equation}
Thus $-\left\langle T^{\tau}_{\tau,e}\right\rangle_E$ can be interpreted as the energy density profile whose form is given by figure \ref{form}(a).
One finds that the energy density turns negative for $\rho < \sqrt{\frac{\alpha'}{2}}$. This implies that the highly excited string matter violates the weak-energy condition. Such behavior was observed in the past for quantum fields near black holes as well \cite{Candelas:1980zt}\cite{Page:1982fm}\cite{Howard:1984qp}\cite{Frolov:1986ut}\cite{Visser:1996ix}\cite{Carlson:2003ub}.\footnote{Note though that there are differences between the results obtained in these papers and our result. In the cited papers, the authors sum the thermal stress tensor and the Casimir stress tensor (both of which are infinite). The sum is finite and has a negative-energy zone for Schwarzschild black holes. For Rindler space on the other hand, the sum is strictly zero \cite{Dowker:1994fi} which equals the vev of the stress tensor in the Minkowski vacuum. In the field theory case, there is a discrepancy between the Rindler result and the Schwarzschild result close to the horizon. It is known that this is related to curvature corrections to the stress tensor vev \cite{Frolov:1989jh}\cite{Zurek:1985gd}. Inspection of our result (the thermal scalar classical stress tensor) for the full black hole (for instance the $SL(2,\mathbb{R})/U(1)$ cigar that we used in \cite{Mertens:2013zya} to arrive at Rindler space) and Rindler space itself shows that for our purposes the results agree: no curvature corrections are needed and Rindler space captures the near-horizon region of the black hole. This is presumably because we only care for the most dominant contribution. For some more recent accounts of the stress tensor near black hole horizons in light of the holographic correspondence, see \cite{Haddad:2013tha}\cite{Figueras:2013jja}.} An obvious difference with field theory is that for string theory the stringy matter becomes of negative energy density at a string scale distance from the black hole (instead of the Schwarzschild scale).\footnote{One can check this explicitly by transforming the string-normalized Rindler space to the Schwarzschild normalization as is done for instance in \cite{Mertens:2013zya}.} With this stringy localization in mind, let us emphasize a point which we skimmed over in \cite{Mertens:2013zya}. In a black hole-normalized geometry
\begin{equation}
ds^2 = \frac{\rho^2}{(4GM)^2}d\tau^2 + d\rho^2 + \hdots,
\end{equation}
the Klein-Gordon type equations for non-winding modes give eigenfunctions that do not depend on the string length $l_s$.\footnote{Their eigenvalues do depend on the string length for massive string states.} The peculiarity of the winding modes is that T-duality explicitly introduces the string length into the eigenvalue problem. This causes the thermal scalar in Rindler space to be localized at string length from the horizon; pure discete momentum modes on the other hand oscillate in space with an oscillation length of the order of the black hole scale $GM$ (\emph{not} the string scale). \\

\noindent Analogously one computes the spatial components of the stress tensor and one finds
\begin{align}
\left\langle T^{\rho\rho}_{e}(\mathbf{x})\right\rangle_E &= \frac{2E}{\sqrt{\alpha'}}e^{-\rho^2/\alpha'}, \\
\left\langle T^{ij}_{e}(\mathbf{x})\right\rangle_E &= \delta_{ij}\frac{2E}{\sqrt{\alpha'}}\left[1- \frac{\rho^2}{\alpha'}\right]e^{-\rho^2/\alpha'},
\end{align}
The radial pressure $\left\langle T^{\rho}_{\rho}(\mathbf{x})\right\rangle$ is given in figure \ref{form}(b) and is found to be positive and localized to the horizon. Curiously, the transverse pressure, depicted in figure \ref{form}(c), changes sign at $\rho=\sqrt{\alpha'}$. An obvious feature is that, since the stress tensor is quadratic in the fields, it decays at twice the rate of $T_0$. Note also that, as discussed above, the highly excited string(s) does not have non-zero net impulse in any direction: $\left\langle T^{\tau \rho}_{e}(\mathbf{x})\right\rangle_E = \left\langle T^{\tau i}_{e}(\mathbf{x})\right\rangle_E = 0$. This agrees with for instance the field theory results for the Hartle-Hawking vacuum of Schwarzschild black holes. The Unruh vacuum for a Schwarzschild black hole on the other hand would include non-zero $T^{\tau \rho}$ corresponding to the flux of particles emitted by a black hole formed in gravitational collapse (Hawking radiation). \\

\noindent Let us discuss some generalities on occurence of negative energy density. Firstly, it is not possible to have $-\left\langle T^{\tau}_{\tau}\right\rangle < 0$ everywhere. This follows immediately from the fact that the total energy is $E$. Secondly, since the spatial kinetic part $G^{ij}\partial_i T \partial_j T^*$ in (\ref{thenergy}) is positive semi-definite, a necessary condition for negative energy density is that 
\begin{equation}
\label{local}
\beta_H^2 G_{\tau\tau} < \beta_{H,\text{flat}}^2.
\end{equation}
This is the condition that the local thermal circle becomes smaller that the flat space Hagedorn circle. For instance for Rindler space, we have that $\rho < \sqrt{2\alpha'}$, and indeed the negative energy density region lies inside this domain. Is satisfying the condition (\ref{local}) uncommon? Actually, all backgrounds must satisfy this condition for some points. The reason is that the thermal scalar is a zero-mode at the Hagedorn temperature (by definition). If this condition is nowhere satisfied, the on-shell action for the thermal scalar (\ref{thaction}) would be positive definite which is impossible.\footnote{An exception occurs if the thermal scalar is space-independent: only then is the spatial kinetic term zero and there is no a priori requirement to have a negative energy density anywhere. Hence this occurs only for spaces with a non-varying thermal circle (meaning constant $G_{\tau\tau}$) since in that case, one can find a constant mode as a solution to the thermal scalar equation. An immediate example of this is flat space. Also for instance 4d Minkowski space combined with a 6d compact unitary CFT (e.g. a compact Calabi-Yau 3-fold) falls in this category.}\\

\noindent For the string charge in Rindler space, we obtain $\left\langle J^{\tau k}_{e}(\mathbf{x})\right\rangle_E = 0$ for all $k$. \\ 

\noindent From a canonical point of view, the situation is analogous: the dominant part of the stress tensor of the near-Hagedorn string gas is given by the classical thermal scalar stress tensor, suitably multiplied by a temperature-dependent factor. Note though that it was previously remarked \cite{Mertens:2013zya}\cite{Sugawara:2012ag} that the Hagedorn temperature equals the Hawking temperature in this case, meaning that in principle the stress tensor becomes infinitely large. The resolution of this infinity might only occur when including higher genus corrections.

\section{Summary}
\label{summary}
Let us summarize the above results.
\begin{itemize}
\item{The thermal scalar energy-momentum tensor captures the energy-momentum tensor of an average of highly excited Lorentzian strings.}
\item{The $U(1)$ charge symmetry of the thermal scalar action leads to a classical Noether current which is the same as the (time-averaged) string charge tensor of an average of highly excited Lorentzian strings.}
\item{One can readily extend these results to the canonical thermal average of the energy-momentum tensor or the string charge. We find expressions with the same spatial distribution. Our main result is equation (\ref{mainresult}) for the energy-momentum tensor in the canonical ensemble.}
\item{A special class of correlators can also be computed using these methods, where each individual stress tensor is time-averaged. The near-Hagedorn dominant behavior (for a fully compact space) is simply the product of the expectation values.}
\item{We have demonstrated these results on three spacetimes. We briefly looked at flat spacetime, demonstrating the presence of a pressureless state of matter at the Hagedorn temperature. $AdS_3$ showed that a non-trivial string charge is possible for backgrounds including Kalb-Ramond fields. Finally we looked at Rindler spacetime, where a negative-energy state of matter was found living close to the event horizon.}
\end{itemize}
In general we conclude that quantum properties of the long string are translated into classical properties of the thermal scalar.

\section*{Acknowledgements}
We thank David Dudal for several discussions. TM thanks the UGent Special Research Fund for financial support. The work of VIZ was partially supported by the RFBR grant 14-02-01185.

\appendix

\section{Heuristic argument why the type II superstring thermal scalar action does not receive $\alpha'$-corrections.}
\label{argu}
We use the following two conjectures on CFTs:
\begin{itemize}
\item{Any rational CFT can be realized as a coset model.}
\item{Any CFT can be arbitrarily well approximated by a rational CFT. A textbook example is the compact scalar which for rational $R^2/\alpha'$ can be rewritten in terms of a larger chiral algebra (and forms a rational CFT).}
\end{itemize}
Thus any SCFT can be approximately written as a coset model. Since the non-winding states have a propagation equation that is simply the Laplacian on the coset manifold, and since both ungauged and gauged WZW models do not receive $\alpha'$ corrections for the type II superstring, all non-winding states simply propagate using the lowest order background fields (Klein-Gordon in curved background).\footnote{Such an argument is not true for bosonic strings: it is known that for instance in WZW $AdS_3$ spacetime (which is an ungauged WZW model), the bosonic string propagation equations have the shift $k\to k-2$ in the Laplacian.} Since this is now for \emph{all} SCFTs, the type II propagation equations in a general background are not $\alpha'$ corrected. Now taking the T-dual equations of motion, also these do not suffer from $\alpha'$ corrections and we conclude that type II superstrings have a thermal scalar equation of motion that coincides with the lowest order in $\alpha'$ effective thermal scalar field equation. \\
We remind the reader that we are only considering terms quadratic in the fields in the action: self-interactions of the thermal scalar will be present in general (even for type II superstrings) but at the non-interacting level we are focusing on in this paper, they are not needed. In particular, the Hagedorn temperature is defined at the one-loop level and hence no self-interactions of the thermal scalar field should be considered. \\
This argument shows that the (non-self-interacting) thermal scalar action for type II superstrings does not get $\alpha'$-corrected. For the derivation presented in section \ref{HP} however, we require more: we need to be able to vary with respect to the background metric. The result is an off-shell background. For this we do not have the CFT methods available. Nonetheless, it seems plausible that a suitable off-shell generalization of string theory (e.g. using double field theory) should not alter this result.

\section{Total energy in a stationary spacetime}
\label{stationary}
\subsection{Local analysis}
Consider an observer in a stationary spacetime moving along a trajectory tangential to $\frac{\partial}{\partial t}$. Hence his 4-velocity equals $u^{\mu} = \frac{1}{\sqrt{-G_{00}}} \frac{\partial}{\partial t}$, where the norm is fixed by requiring $u^{\mu}u_{\mu} = -1$. The energy density as measured by such an observer equals $T^{\mu\nu}u_{\mu}u_{\nu}$. In writing these expressions we are assuming that this vector is globally timelike in the globally hyperbolic section of the spacetime onto which we are focusing; this excludes for instance generic Kerr-Newman black holes. Note though that there exist rotating black holes which do have globally timelike Killing vectors (such as a subclass of Kerr-AdS black holes \cite{Hawking:1999dp}\cite{Winstanley:2001nx}). Before continuing, we make the transition to the Euclidean signature manifold. This (Euclidean) energy density can be expanded as
\begin{equation}
\label{total}
T^{\mu\nu}u_{\mu}u_{\nu} = T^{\tau\tau}G_{\tau\tau} + 2 T^{\tau i} G_{i\tau} + T^{ij}\frac{G_{i\tau}G_{j\tau}}{G_{\tau\tau}}.
\end{equation}
and it is related to the Lorentzian energy density by a sign change. \\

\noindent First we compute only $T^{\tau\tau}G_{\tau\tau} + T^{\tau i} G_{i\tau} = T^{\tau}_{\tau}$. We will show that this is precisely the total energy. Thus we consider
\begin{align}
-\int \left\langle T_{\tau,e}^{\tau}\right\rangle_E \sqrt{G} dV &= -\int \left(G_{\tau\tau}\left\langle T^{\tau\tau}_{e}\right\rangle_E + G_{\tau k}\left\langle T^{\tau k}_{e}\right\rangle_E\right) \sqrt{G} dV.
\end{align}
This expression has three different contributions. Part of the first term is the same as in the static case and yields the total energy:
\begin{align}
\frac{E}{\beta_H^{2}} \frac{\int dV \sqrt{G} G_{\tau\tau}\frac{\beta_H^2 TT^*}{4\pi^2\alpha'^2} }{\left.\frac{\partial \lambda_0}{\partial \beta^2}\right|_{\beta = \beta_H}} = E. 
\end{align}
For the other terms we also need
\begin{align}
\frac{e^{2\Phi}}{\sqrt{G}}\frac{\delta \lambda_0}{\delta G_{\tau i}} &= -\left( G^{ki} G^{\tau l} + (k \leftrightarrow l)\right)\left(\frac{\partial_k T^* \partial_{l} T + \partial_{l}T^* \partial_{k} T}{4}\right) \nonumber \\
&+ \frac{1}{2}G^{\tau i}\left[G^{kl}\partial_k T \partial_l T^* + \frac{\beta^2G_{\tau\tau}-\beta_{H,\text{flat}}^2}{4\pi^2\alpha'^2}TT^*\right].
\end{align}
A second term (proportional to the Lagrangian), vanishes after using $G_{\tau\tau}G^{\tau\tau} + G_{\tau i}G^{\tau i} = 1$ due to the fact that the thermal scalar is a zero-mode. \\
Finally, the remaining term equals
\begin{align}
\frac{E}{\beta_H^{2}} \frac{\int dV \sqrt{G} \left[ - G_{\tau\tau}G^{\tau i}G^{\tau j}\partial_i T \partial_j T^* - \frac{G_{\tau k}}{2}\left(G^{ki}G^{\tau j} + G^{j k}G^{\tau i}\right)\partial_i T \partial_j T^*\right]}{\left.\frac{\partial \lambda_0}{\partial \beta^2}\right|_{\beta = \beta_H}}.
\end{align}
Upon using $G_{\tau k}G^{i k}+ G_{\tau\tau}G^{\tau i} =0$, one readily finds this term vanishes as well. In all, one finds that $-\left\langle T_{\tau,e}^{\tau}\right\rangle_E$ represents the energy density. \\

\noindent Next we show that the remainder of (\ref{total}) vanishes. One readily obtains
\begin{equation}
\frac{\delta \bar{G}^{ij}}{\delta G_{kl}} = - \frac{G^{ik} G^{jl} + (i \leftrightarrow j)}{2}.
\end{equation}
Using this result we obtain
\begin{align}
\frac{e^{2\Phi}}{\sqrt{G}}\frac{\delta \lambda_0}{\delta G_{kl}} &= -\left( G^{ik} G^{j l} + (i \leftrightarrow j)\right)\left(\frac{\partial_i T^* \partial_{j} T + \partial_{j}T^* \partial_{i} T}{4}\right) \nonumber \\
&+ \frac{1}{2}G^{kl}\left[G^{ij}\partial_i T \partial_j T^* + \frac{\beta^2G_{\tau\tau}-\beta_{H,\text{flat}}^2}{4\pi^2\alpha'^2}TT^*\right].
\end{align}
The term on the second line of this expression combines with the second term of $T^{\tau i} G_{i\tau}$, with the factor in front of the square brackets yielding
\begin{equation}
G_{\tau i}G^{\tau i} + G^{kl}\frac{G_{k\tau}G_{l\tau}}{G_{\tau\tau}} = G_{\tau i}G^{\tau i} - G^{\tau k}G_{\tau k} = 0.
\end{equation}
The other terms combine into
\begin{align}
\frac{E}{\beta_H^{2}} \frac{\int dV \sqrt{G} \left[ - \frac{G_{\tau k}}{2}\left(G^{ki}G^{\tau j} + G^{j k}G^{\tau i}\right)\partial_i T \partial_j T^* - \frac{G_{\tau k}G_{\tau l}}{2G_{\tau\tau}}\left(G^{ik}G^{jl} + G^{j k}G^{il}\right)\partial_i T \partial_j T^*\right]}{\left.\frac{\partial \lambda_0}{\partial \beta^2}\right|_{\beta = \beta_H}}.
\end{align}
The prefactors (for fixed $i$ and $j$) can be seen to vanish by again using $G_{\tau l}G^{jl} + G_{\tau\tau}G^{\tau j} = 0$.

\subsection{Komar integral}
\begin{figure}[h]
\centering
\includegraphics[width=0.5\textwidth]{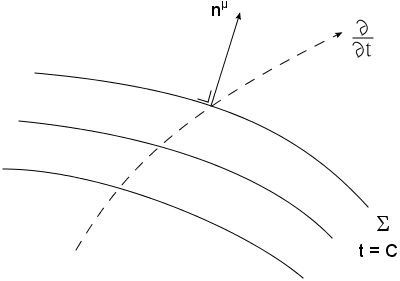}
\caption{Constant time hypersurfaces and their normal 4-vector. Also shown is the flow of the timelike Killing vector. For a non-static spacetime, these vectors are not parallel.}
\label{curvilinear}
\end{figure}

There exists an alternative method of determining the total energy according to the asymptotic observer, the Komar integral:
\begin{equation}
\label{Kom}
E= \int_{\Sigma} T^{\mu\nu}\xi_{\mu}n_{\nu} d\Sigma,
\end{equation}
where we integrate over a spacelike hypersurface with normal $n^{\mu}$ and where $\xi^{\mu}$ is the timelike Killing vector. For static spacetimes it is readily seen that this definition coincides with the description used above, but for stationary spacetimes this seems not immediate. Let us consider this case in detail. A defining feature of non-static (but stationary) spacetimes is the fact that the normal vector to spatial slices is not parallel to the tangent of the timelike Killing vector. This implies that one can define two different notions of being constant in time. A sketch of this situation is given in figure \ref{curvilinear}. 
The 4-vector $n^{\mu}$ is normal to the constant time hypersurfaces and is hence of the form:
\begin{equation}
n^{\mu} = \frac{\nabla^{\mu} \tau}{\left(\nabla^{\nu} \tau \nabla_{\nu} \tau\right)^{1/2}} = \frac{G^{\tau\mu}}{\sqrt{G^{\tau\tau}}}.
\end{equation}
Its covariant components are readily found to be
\begin{align}
n_\tau &= \frac{G_{\tau\tau}G^{\tau\tau}}{\sqrt{G^{\tau\tau}}} + \frac{G_{\tau i}G^{\tau i}}{\sqrt{G^{\tau\tau}}} = \frac{1}{\sqrt{G^{\tau\tau}}}, \\
n_i &= \frac{G_{i\tau}G^{\tau\tau}}{\sqrt{G^{\tau\tau}}} + \frac{G_{i j}G^{\tau j}}{\sqrt{G^{\tau\tau}}} = 0.
\end{align}
The other ingredient is the timelike Killing vector. Its covariant components are
\begin{align}
\xi_\tau &= G_{\tau\tau}, \\
\xi_i &= G_{\tau i}.
\end{align}
The spacelike slice $\Sigma$ is integrated over with the pull-back metric $G_{ij}$ (setting $dt=0$ in the full metric). \\
With these formulas, the integrand of (\ref{Kom}) equals
\begin{align}
\sqrt{G_{ij}} T^{\mu\nu}\xi_{\mu}n_{\nu} = \frac{\sqrt{G_{ij}}}{\sqrt{G^{\tau\tau}}}\left[ T^{\tau\tau}G_{\tau\tau} + T^{\tau i}G_{\tau i}\right].
\end{align}
To proceed, we need the following identities:
\begin{align}
\sqrt{G} = \sqrt{G_{ij}}\sqrt{G_{\tau\tau} - G_{\tau k}G_{\tau l}\bar{G}^{kl}}, \\
G^{\tau\tau}\left(G_{\tau\tau} - G_{\tau k}G_{\tau l}\bar{G}^{kl}\right) = 1.
\end{align}
In these formulas, we have written 
\begin{equation}
\bar{G}^{kl} = G^{kl} - \frac{G^{\tau k}G^{\tau l}}{G^{\tau\tau}},
\end{equation}
which is the matrix inverse of the purely spatial matrix $G_{ij}$. \\

\noindent We hence have 
\begin{align}
\sqrt{G_{ij}} T^{\mu\nu}\xi_{\mu}n_{\nu} = \sqrt{G} T^{\tau}_{\tau},
\end{align}
precisely the same as before. \\

\noindent The physical interpretation of the previous result is that we should redshift the local energy density $T_{\mu\nu}u^{\mu}u^{\nu}$ on the spatial slices $\Sigma$. For stationary spacetimes, the redshift factor is $\sqrt{G_{\tau\tau}}$ which combines with the spatial metric as defined by observers moving along $\frac{\partial}{\partial t}$ to yield the total background metric:
\begin{equation}
\sqrt{G} = \sqrt{G_{\tau\tau}}\sqrt{G_{ij}-\frac{G_{\tau i}G_{\tau j}}{G_{\tau\tau}}}.
\end{equation}
We conclude that also for stationary spacetimes we have that
\begin{equation}
\boxed{
E = \int_{\Sigma} T^{\mu\nu}\xi_{\mu}n_{\nu} \sqrt{G_{ij}} d\Sigma = \int_{\Sigma} T^{\mu\nu}u_{\mu}u_{\nu} \sqrt{G} d\Sigma}.
\end{equation}
The first equality is the global Komar analysis, whereas the second expression is interpreted as summing (and redshifting) the local energy density over the entire spatial slice using the spatial metric as constructed by the radar definition of the stationary observers. \\
This story applies to string theory, in spite of the fact that the dominant near-Hagedorn behavior implies non-local long strings. The crucial point is that this long string is effectively treated as a Euclidean quantum field whose energy density can be viewed as local in space.

\end{document}